\documentclass[12pt]{article}
\usepackage{amsmath,amssymb}
\usepackage{eucal}

\usepackage[dviwindo]{graphicx}

\newcommand{\bs}{\begin{subequations}}
\newcommand{\es}{\end{subequations}}
\numberwithin{equation}{section}
\def \myfigures #1#2#3#4
{\begin{figure}[ht]
    \begin{center}
        \includegraphics[width=#2 \textwidth]{#1.eps}
        \hfill
        \includegraphics[width=#4 \textwidth]{#3.eps}
    \end{center}
\end{figure}}

\newcommand{\ben}{\begin{eqnarray}}
\newcommand{\een}{\end{eqnarray}}
\newcommand{\la}{\label}
\begin{document}

\title{Static Fundamental Solutions of Einstein Equations and
Superposition Principle in Relativistic Gravity}

\vskip 1.5truecm

\author{P.~P.~Fiziev\thanks{Department of Theoretical Physics, University of
Sofia, Boulevard 5 James Bourchier, Sofia 1164, Bulgaria
E-mail:\,\,\,fiziev@phys.uni-sofia.bg}}

\date{}
\maketitle

\begin{abstract}
We show that Einstein equations are compatible with the presence of
massive point particle idealization and find the corresponding two
parameter family of solutions. They are complete defined by the bare
mechanical mass $M>0$ and the Keplerian mass $m>0$ ($m < M$) of the
point source of gravity. The global analytical properties of these
solutions in the complex plane define a unique preferable radial
variable of the one particle problem.

These new solutions are fundamental solutions of the quasi-linear
Einstein equations. We introduce and discuss a novel nonlinear
superposition principle for solutions of Einstein equations and
discover the basic role of the relativistic analog of the Newton
gravitational potential. For the relativistic potential we introduce
a simple quasi-linear superposition principle as a new physical
requirement for the initial conditions for Einstein equations, thus
justifying the instant gravistatic case for N particle system.

This superposition principle allows us to sketch a new theory of the
gravitational mass defect. In it a specific Mach-like principle for
the Keplerian mass $m$ is valid, i.e. it depends on the mass
distribution in the universe, in contrast to the bare mass $M$,
which remains a true constant. Several basic examples both of
discrete and of continuous mass distributions are considered.
\end{abstract}

\sloppy

\section{Introduction}
\subsection{Static Fundamental Solution and Superposition Principle in Newton Theory of Gravity}
The notion of a static {\em fundamental solution} of a classical
field equation appeared at first in Newton theory of gravity
\cite{books}. Such a solution  solves the Poisson equation with
source term, proportional to Dirac 3D $\delta$-function:
\ben \Delta \varphi^{{}^{{}_{Newton}}}({\bf r})=4\pi
m\delta^{(3)}({\bf r}). \la{Poisson}\een
Here and further on for simplification of the formulas we are
using units in which the Newton gravitational constant
$G^{{}^{{}_{Newton}}}=1$ and velocity of light $c=1$.

The static fundamental solution does not depend on arbitrary
functions, or additional constants. It is unambiguously fixed, among
all solutions of Eq. (\ref{Poisson}), if we require this solution to
tend to zero at infinite distances. This unique solution describes
the potential of the newtonian gravitational field
\ben \varphi^{{}^{{}_{Newton}}}({\bf r})=-{{m}\over r},
\la{FSN}\een
which is created by a classical point particle of gravitational
(Keplerian) mass $m$, placed in Euclidean 3D space at the origin
of coordinate system ${\bf{r}}_0={\bf 0}$. Here $r=|{\bf r}|\geq
0$.

Analogous solutions are well known in the problem of static point
source of electric field in Maxwell electrodynamics,  as well. A
proper generalization of the notion of fundamental solution for
hyperbolic partial differential equations can be find, for
example, in \cite{YCB} and in the references therein.

According to well known mathematical results, the solutions
(\ref{FSN}) describe, too, the static field in vacuum, outside
sources of finite dimension, assuming spherical symmetry of the
corresponding distribution $\mu({\bf r})=\mu(r)$  of gravitational
mass, or electric charge.

The fundamental role of the solutions (\ref{FSN}) in Newton
gravistatics is substantiate by the {\em superposition principle},
according to which in these linear theories the field of any
aggregate of matter can be obtained as a {\em sum} of the fields
of its constituent matter points at positions ${\bf r}_A$ (or
${\bf r^\prime}$). For example, in Newton gravistatics
\ben \varphi^{{}^{{}_{Newton}}}({\bf r; r_1,\dots,r_N
})=-\sum_{A=1}^N{{m_A}\over {|{\bf r-r_A}|  } } \la{DSPP},\een
in the case of a set of $N$ discrete massive points, and
\ben \varphi^{{}^{{}_{Newton}}}({\bf r})=-\int
{{\mu_{{}_{Kepler}}({\bf r^\prime})}\over {|{\bf r-r^\prime}|  } }
d^3{\bf r^\prime} \la{SPP}\een
in the case of a continuous distribution of Keplerian mass
$\mu_{{}_{Kepler}}({\bf r})$.

In a more general form the superposition principle in the newtonian
gravity can be expressed as

{\bf Proposition 1:}
{\em If $\mu_{I}({\bf r})$ and $\mu_{II}({\bf r})$ are two mass
distributions, which create gravitational fields with corresponding
potentials $\varphi^{{}^{{}_{Newton}}}_{I}({\bf r})$ and
$\varphi^{{}^{{}_{Newton}}}_{II}({\bf r})$, then the potential of
the field, created by mass distribution $\mu({\bf r})=\mu_{I}({\bf
r})+\mu_{II}({\bf r})$ is}
\ben \varphi^{{}^{{}_{Newton}}}({\bf
r})=\varphi^{{}^{{}_{Newton}}}_{I}({\bf
r})+\varphi^{{}^{{}_{Newton}}}_{II}({\bf r}).\la{SPPI}\een

\subsection{Massive Point Particle in General Relativity}

\subsubsection{The Schwarzschild Solution}

An attempt to solve the point-mass problem in general relativity
(GR) was made at first ninety years ago, as early as in the
pioneering article by Schwarzschild  \cite{Schwarzschild} and its
subsequent modifications \cite{HDW, Kruskal}.

Today it is well known that inconsistences arise when we look at
Schwarzschild solution as the space-time arising from localized
point mass singularity \cite{NP}. Actually, the well known
Schwarzschild metric in Hilbert gauge:

\ben ds^2\!=\left(\!1-{{2m}\over\rho}\right)dt^2+{{dr^2}\over
{1-{{2m}\over\rho}}}
 -\rho^2(d\theta^2+\sin^2\theta\,d\phi^2)
 \la{Hilbert}\een
solves the {\em vacuum} Einstein equations $G_\mu^\nu=0$ in the
spherically symmetric static case. It possesses an {\em event
horizon} at $\rho=\rho_G=2m$ and  a strong hidden {\em singularity}
at $\rho=0$. This solution describes a completely empty space-time
with removed point $\rho=0$ and nontrivial non-Euclidean topology.
Indeed, in the Weyl's isotropic coordinates with radial variable
$r_W={\frac 1 2}\left(\rho-m+\sqrt{\rho(\rho-2m)}\right)$ one can
easily see that this solution describes a two sheeted space-time
with a flat asymptotic at $r_W=0$ and $r_W=\infty$ connected by a
specific bridge. Then the nonzero Keplerian mass $m$ appears in the
solution due to the nontrivial topology of the space-time, i.e. in
the spirit of Einstein-Rosen-Misner-Wheeler geometrodynamics
\cite{Wheeler}, as described by the sentence "mass, without mass".

The singularity at $\rho=0$ {\em is not related} with a massive
point particle with proper bare mass $M$ and mechanical action
${\cal A}_M\!=\!-M\int\!ds$. Indeed, as we shall see in more detail
below, the solution (\ref{Hilbert}) does not solve the Einstein
equations (EE)
\ben G^\mu_\nu = 8\pi T^\mu_\nu \la{EE}\een
in presence of matter with stress-energy tensor $T^\mu_\nu \sim
M\,\delta^{(3)}({\bf r})$. Here $M\delta^{(3)}({\bf r})$ describes
the mass distribution of the point particle with proper bare mass
$M$.

In his pioneering article Schwarzschild has used another radial
variable $r$, which defers essentially from variable $\rho$, i.e.,
his choice a radial gauge for the spherically symmetric static
metric
\ben
ds^2\!=\!g_{tt}(r)^2\,dt^2\!+\!g_{rr}(r)^2\,dr^2\!-\!\rho(r)^2(d\theta^2\!+
\!\sin^2\theta\,d\phi^2) \la{4Dinterval}\een
is different from the Hilbert's one (\ref{Hilbert}).

Borrowing from the Minkowskian flat space-time the gauge condition
$$|{}^4\!g|:=\det||g_{\mu\nu}(t,{\bf r})||=1$$ (which takes place
there in Cartesian coordinates), Schwarzschild  was able to fix
the three unknown functions in the form:
$$\rho(r)=\sqrt[3]{r^3+\rho_G^3}>0,\,\,\,
g_{tt}(r)=1-{{2m}\over\rho(r)}>0, \,\,\,\hbox{and}\,\,\,
g_{rr}(r)=-\rho^\prime(r)^2/g_{tt}(r)<0.$$

This solution of EE has {\em no event horizon}. Its peculiar feature
is that it describes a {\em point like object} of Keplerian mass
$m>0$, zero radius, zero volume, but {\em nonzero} area
$A_{\rho_G}=4\pi\rho_G^2>0$. These unusual properties of the {\em
original} Schwarzschild solution have been discussed by Brillouin
\cite{Brillouin} as early as in 1923.

At present the original Schwarzschild geometry and other similar
geometries of space-time are widely ignored in GR. A main stream of
articles in the last 40 years is strongly limited to consideration
of the black hole interpretation of the Hilbert form (\ref{Hilbert})
of vacuum Schwarzschild solution and its generalizations. In
addition, essential features of the original Schwarzschild solution
are not reproduced in the most of modern literature on this subject
and remain hardly known.

\subsubsection{On the Choice of Radial Variable in the  GR Massive Point
Particle Problem}

One of the basic problems in the description of single massive
point particle as a source of gravitational field in GR is the
choice of proper radial variable $r$.

The quantity $\rho\geq 0$ has a clear geometrical and physical
meaning:

i) It is well known that $\rho$ defines the area $A_\rho=4\pi\rho^2$
of a centered at $r=0$ sphere with "area radius" $\rho$ and the
length of the big circle on it $l_\rho=2\pi\rho$. Thus we see that
the quantity $\rho$ has a well defined geometrical meaning and is a
gauge invariant notion.

ii) The coordinate $\rho$ measures the curvature $\sim 1/\rho^2$
of the 2D-manifolds (2D-spheres) in 3D Riemannian space, which are
invariant under rotations around the center of spherical symmetry.
It measures, too, the curvature of the 4D pseudo-Riemannian
space-time: ${{}^{4}}\!R={{}^{4}}\!R(\rho)$ and of the
corresponding 3D-space: ${{}^{3}}\!R={{}^{3}}\!R(\rho)$ in the
spherically symmetric case, inside the matter source of finite
dimension. Hence the name "curvature radius".

iii) From physical point of view one can refer to $\rho$ as an
optical "{luminosity distance}", because the luminosity $L$ of
distant physical objects is reciprocal to $A_\rho$: $L\sim
{1/{\rho^2}}$.

In contrast, the {physical and geometrical meaning} of the
coordinate $r$  is not defined by the spherical symmetry of the
problem and is unknown {\em a priori} \cite{Synge}. Its choice has
been discussed from physical point of view by Eddington as early
as in \cite{Eddington}. His conclusion was that all admissible
variables $r$ are practically equivalent at distances
$r>>\rho_{G}$, since under suitable coherent choice of their
scales we have $\rho/r\to 1$ when $r\to\infty$.

The following assumptions about the mathematical properties of the
radial variable $r$ of the single point particle problem seem to
be natural from physical point of view:

i) Its value $r=0$ is to correspond to the center of the symmetry,
where one must place the physical source of the gravitational
field -- the massive point particle.

ii) The radial variable is to vary in the semi-bounded physical
interval $r\in [0,\infty)$.

iii) The luminosity variable $\rho(r)$ is to increase monotonically
to infinity in this interval, together with radial ones, i.e.,
$d\rho/dr>0$, and, in addition, one has to impose the Eddington
condition:
\ben \lim\limits_{r\to\infty}{{\rho(r)}\over r}=1.
\la{rho_r_limit}\een

iv) The infinite value $r=\infty$ of the radial variable is to be
prescribed to the boundary of the asymptotically flat domain of
space-time.

v) There must not exist non-physical singularities of the solution
of EE in the whole compactified {\em complex} domain
$\mathbb{\widetilde{C}}$ of the radial variable $r \in
\mathbb{\widetilde{C}}$.

In the present article we show that these physical requirements
define in a unique way the radial variable $r$ of the problem at
hand, thus solving the corresponding uniformization problem in the
case of point particle source of gravity in GR.

\subsubsection{Some Remarks on the Massive Point Particle Idealization in GR}

A clear physical motivation for consideration of massive point
particle sources of gravitational field in GR, both electrically
neutral and charged ones, can be found in 1962-63 Feynman lectures
on gravity \cite{Feynman}. The energy momentum tensor of a point
particle has been used in the excellent textbooks by Landau \&
Lifschiz and by Weynberg \cite{books} as a tool for treatment of
many particle systems in GR. In spite of this fact the single
particle case is still an open problem in GR.

Moreover, at present the vast majority of relativists do not accept
the consideration of point particles in GR, assuming that it is an
idealization, which is incompatible with EE  \cite{MacCallum}. There
are different reasons:

i) Some doubts about consistence of the theory of mathematical
distributions (like 3D Dirac $\delta$-function $\delta^{(3)}({\bf
r})$) \cite{Gelfand} with the obviously nonlinear character of EE
\cite{YB}.

The formal mathematical problems, which emerge when one attempts to
work with distributions in EE were successfully advanced in the last
decade using Colombeau's theory of generalized functions
\cite{Colombeau_GR}. Unfortunately, the published results on the
point particle problem in GR, based on this approach, are physically
incorrect (see Section 2.3.5).

ii) The clear understanding  that an infinite concentration of
energy in a single space-point will change drastically the
geometry of the GR-Riemannian space-time
$\mathbb{M}^{(1,3)}\{g_{\mu\nu}\}$ in a small vicinity of the
world-line of this point;

iii) Some attempts to neglect the role of classical description of
matter in GR, replacing it by classical field description, or by
quantum field description, according to the so called "third
approach" by Einstein, Wheeler and many others, see \cite{Wheeler}
and references therein;

iv) The absence of understanding of necessity to use fundamental
solutions of EE. These were unknown up to recently in GR, but may
turn to be useful mathematical tool.

v) The absence of a general non-linear superposition principle for
EE, which is to correspond to the linear superposition principle
(\ref{DSPP}) -- (\ref{SPPI}) in Newton theory of gravity. Note that
a specific kind of superposition principle for initial conditions of
black hole solutions is well known \cite{Wheeler}, but the general
problem and other specific cases are still not studied, to the best
of our knowledge.

On the other hand, it is obvious that in Nature the very distant
stars look like "points" of finite mass and finite luminosity. This
fact has a proper mathematical description in the language of
mathematical distributions in the Newton theory of gravity, but
still not in GR.

In spite of the absence of proper description of the massive point
particles, in the practical relativistic celestial mechanics, for
example, in the calculations of the solar-system trajectories of the
space crafts, even the Sun is considered often as a massive point
source of gravity.

A formal mathematical problem is to find the corresponding correct
treatment of such objects in GR, but up to recently no reasonable
approach was known. Unfortunately, the most of the existing formal
attempts to solve the point particle problem in GR do not take into
account one essential {\em physical} difference between the GR and
the Newtonian description of the massive objects. It is well known
\cite{books}, that any body in GR has two different masses: the
Kepelrian one $m$, as seen from distant observer, and the proper
(bare) mass $M>m$, which is the sum of the masses of its
constituents, when placed at infinite distances between them, i.e.,
with gravitational interaction -- turned off. The difference $M-m$,
or the ratio $\varrho=m/M$  describe the gravitational defect of
mass. One must include properly this specific feature of the
relativistic theory of gravity in the GR-point-particle model. To
the best of our knowledge, such attempts ware not made up to
recently, with the only exception -- \cite{PF}.

In the present article we show that a correct mathematical solutions
of EE with $\delta^{(3)}({\bf r})$ term in the rhs do exist. Such
solutions describe a two parameter family of analytical space-times
$\mathbb{M}^{(1,3)}\{g_{\mu\nu}\}$ with a specific strong
singularity at the place of the massive point source with bare
mechanical mass $M>0$ and Keplerian mass $m<M$.

The price, one has to pay for this enlargement of the standard GR
framework, is:

i) To accept the unusual geometry of the space-time around the
matter point with infinite concentration of energy in it.

As we have stressed already, this geometry was introduced in GR for
the first time  actually in the original Schwarzschild article
\cite{Schwarzschild} and has been discussed by Brillouin
\cite{Brillouin}. The unusual geometry is essentially different from
the geometry around the space-time points with finite energy density
in them.

The global properties of the space-time manifolds, generated by
massive point source, are essentially different from the ones of the
most popular, at present, Hilbert form (\ref{Hilbert}) of the
original Schwarzschild solution, which describes an empty space-time
with nontrivial topology in the spirit of the Einstein-Rosen-Wheeler
geometrodynamics \cite{Wheeler}.

ii) To allow consideration of metrics, whose coefficients are not a
${\cal C}^3$ - smooth functions. Indeed, to reproduce the
$\delta^{(3)}({\bf r})$ term in the rhs of EE, the metric tensor,
and/or its derivatives, related to the geometry of the Riemannian
space-time, must have a definite singularities (discontinuities) at
the place of the point source of gravity \cite{PF}.

iii) To replace the mathematical theory of the real smooth
manifolds, which is in current use in GR, with the theory of the
analytical manifolds with proper singular points, considering the
whole complex domain of the space-time variables.

\subsubsection{The Nonlinear Superposition Principle in GR}

In addition, here we formulate a GR-{\em nonlinear} superposition
principle, analogous to the Newton one, described in Proposition 1.

Let's consider for simplicity only the case of asymptotically flat
space times, which correspond to energy-momentum stress tensors
$T^\mu_\nu (x)$ with compact support in
$\mathbb{M}^{(1,3)}\{g_{\mu\nu}\}$, i.e. let's focus our attention
on so called "island universes". It is well known that after a
proper fixing of the gauge, the boundary conditions at infinity
define the solutions of EE (\ref{EE}) in a unique way, see for
example \cite{YB, books} and the references therein.

{\bf Proposition 2:} {\em Let ${g_{\mu\nu}(x)}_{I}$ and
${g_{\mu\nu}(x)}_{II}$ are two metrics, which correspond via EE
(\ref{EE}) to two energy-momentum tensor distributions ${T^\mu_\nu
(x)}_{I}$ and ${T^\mu_\nu (x)}_{II}$ of compact supports. Then the
metric $g_{\mu\nu}(x)$ of the GR gravitational field, created by
energy-momentum tensor distribution ${T^\mu_\nu (x)}={T^\mu_\nu
(x)}_{I}+{T^\mu_\nu (x)}_{II}$ via EE (\ref{EE}) is uniquely
defined by the two metrics ${g_{\mu\nu}(x)}_{I}$ and
${g_{\mu\nu}(x)}_{II}$.}

Thus we obtain the unambiguous correspondence
\ben \{ {g_{\mu\nu}(x)}_{I} , {g_{\mu\nu}(x)}_{II}\} \mapsto
g_{\mu\nu}(x)\la{RSPPI}\een
and the metric $g_{\mu\nu}(x)$ deserves to be called a nonlinear
superposition of the metrics ${g_{\mu\nu}(x)}_{I}$ and
${g_{\mu\nu}(x)}_{II}$.

The essence of the proof of the existence of such nonlinear
superposition principle in GR is in the simple note that the support
of the distribution ${T^\mu_\nu (x)}={T^\mu_\nu (x)}_{I}+{T^\mu_\nu
(x)}_{II}$ will be certainly compact, if both ${T^\mu_\nu (x)}_{I}$
and ${T^\mu_\nu (x)}_{II}$ have compact supports in
$\mathbb{M}^{(1,3)}\{g_{\mu\nu}\}$. Then the EE (\ref{EE}) with
${T^\mu_\nu (x)}$ in rhs, supplied with asymptotically flat
space-time boundary conditions, will have an unique solution
$g_{\mu\nu}(x)$, which corresponds to the metrics
${g_{\mu\nu}(x)}_{I}$ and ${g_{\mu\nu}(x)}_{II}$ and ought to be
named their (nonlinear) superposition due to obvious physical
reasons. It is clear that $g_{\mu\nu}(x)$ is a very complicated
functional of the metrics ${g_{\mu\nu}(x)}_{I}$ and
${g_{\mu\nu}(x)}_{II}$. The problem of reconstruction of
$g_{\mu\nu}(x)$ in (\ref{RSPPI}), using two given metrics
${g_{\mu\nu}(x)}_{I}$ and ${g_{\mu\nu}(x)}_{II}$ is highly
nontrivial. We shall use the symbol $\circledS$ to denote the
composition (\ref{RSPPI}) of the two metrics in the form
\ben g_{\mu\nu}(x)={g_{\mu\nu}(x)}_{I}\,\circledS\,
{g_{\mu\nu}(x)}_{II}.
 \la{S}\een
It is clear that by construction this new operation on the metrics
is symmetric and associative:
\ben {g_{\mu\nu}(x)}_{I}\,\circledS\,
{g_{\mu\nu}(x)}_{II}={g_{\mu\nu}(x)}_{II}\,\circledS\,
{g_{\mu\nu}(x)}_{I}, \nonumber\\
\Big({g_{\mu\nu}(x)}_{I}\,\circledS\,
{g_{\mu\nu}(x)}_{II}\Big)\,\circledS\,
{g_{\mu\nu}(x)}_{III}={g_{\mu\nu}(x)}_{II}\,\circledS\,
\Big({g_{\mu\nu}(x)}_{I} \,\circledS\,
{g_{\mu\nu}(x)}_{III}\Big).\la{Ssym}\een
These properties are an immediate consequences of the corresponding
properties of the summation of energy-momentum tensors of compact
supports in $\mathbb{M}^{(1,3)}$, assuming that we are considering
space-times with a fixed flat-geometry-boundary-conditions at space
infinity.

Finally, we can define correctly the superposition of an arbitrary
number $N$ of metrics, obeying the same boundary conditions at
infinity, i.e. we can introduce a multiple superposition
operation:
\ben g_{\mu\nu}(x)=\circledS_{A=I}^N\,
{g_{\mu\nu}(x)}_{A}:={g_{\mu\nu}(x)}_{I}\,\circledS\,\dots
\,\circledS\,{g_{\mu\nu}(x)}_{N}.\la{SN} \een

To some extend the novel principle (\ref{S}), (\ref{SN}) is
unexpected, and certainly much more complicated than the simple
linear superposition principle (\ref{SPPI}) in Newton gravity.

Indeed, in GR we have a very specific physical situation. It is
clear that even if the metrics ${g_{\mu\nu}(x)}_{I}$ and
${g_{\mu\nu}(x)}_{II}$ are static, in general case their
superposition $g_{\mu\nu}(x)$ is not a static metric. It contains
the whole GR dynamics, including the possible radiation of
gravitational waves, due to the gravitational interaction between
the physical sources of the metrics ${g_{\mu\nu}(x)}_{I}$ and
${g_{\mu\nu}(x)}_{II}$. In contrast to the situation in
electrodynamics, where we can introduce non-electrodynamical forces
and stresses to keep the composite aggregate of charges in a static
state without introducing new terms in the Maxwell equations, in GR
any additional interactions, introduced for the same purpose, will
have a nonzero energy-momentum tensor ${T^\mu_\nu
(x)}_{additional}$, which enters the rhs of EE and changes the
space-time geometry and the very problem. As a result we see that
without introducing of non-gravitational interactions between
particles in GR we have only a unique (whole-time) static case --
the single point particle problem (see, for example, Fock in
\cite{books}).

Nevertheless, as we shall show in the present article, considering
just {\em the instant} static case, one can introduce a simple
quasi-linear superposition principle for static fundamental
solutions in GR. It reveals the role of static fundamental solutions
of EE in GR, which is much like the role of corresponding
fundamental solutions in linear field theories like Newton gravity
and Maxwell electrostatics, if one considers only a single 3D
space-time surface $t=0$ \cite{Misner}.

The resolution of the relativistic gravistatic problem requires the
solution only of the well known suitable form of $tt$\,-\,EE. It
does not contain second derivatives of metric with respect to the
time variable $t$. Let us consider a space-time, which is a solution
of EE. The 3D curvature ${}^{(3)}\!R$ of arbitrary 3D space-like
surface in it obeys the well known basic equation
\ben {}^{(3)}\!R+K_2=16\pi\mu.\la{RCEK}\een
Here $K_2=K^2-K_{ij}K^{ij}$ (where $K=g^{ij}K_{ij}$) is the
exterior curvature of the 3D surface and $\mu$ is the relativistic
density of mass distribution.

If the last equation is fulfilled at some time instant $t$, as a
consequence of the EE it will be fulfilled at all time instants
$t\in (-\infty,\infty)$, for which the problem is well defined.
There exists an inverse theorem, too: The whole system of EE, which
governs dynamics in GR, may be derived by the requirement to have
the above relation (\ref{RCEK}) co-variantly valid at all time
instants \cite{Foures}.

According to articles \cite{Misner}, one can define the relativistic
gravistatics as a description of time-symmetric initial value
problem for EE in proper coordinates, as well. To see this, it is
enough to know that choosing appropriate coordinates {\em outside}
the 3D surface $t=0$ one obtains for the coefficients of the second
fundamental form of this surface $K_{ij}=-{1\over 2}\partial_t
g_{ij}$. Then one defines the instantaneous-static solutions of EE,
possessing a 3D space-like surface $t=0$ on which $K_{ij}=0$. The
coordinate independent way to this definition implies existence of
isometry of space-time: $t\to -t, {\bf r}\to{\bf r}$ \cite{Misner}.
Then the basic equation reduces to the following simple form, valid
at time instant $t=0$:
\ben {}^{(3)}\!R=16\pi\mu. \la{3R_eq}\een

This equation is not a dynamical equation, but just a constraint on
the initial conditions -- a specific relativistic constraint
equation (RCE). As a result of relativistic dynamics, governed by
the other EE, the RCE will be automatically fulfilled for any time
$t$, if it will be valid at time instant $t=0$ \cite{Misner, books}.
Hence, the time $t$ is a simple auxiliary parameter in the RCE {\em
and} in its solutions. Thus we see that it is enough to solve the
RCE only at the initial time instant $t=0$, i.e., it is enough to
solve the Eq. (\ref{3R_eq}).

From pure mathematical point of view any of the solutions of Eq.
(\ref{3R_eq}) may be considered as an initial condition of a proper
initial value problem for EE. One of the basic purposes of present
article is to find the {\em physically} meaningful solutions of RCE
among the whole variety of its possible mathematical solutions.
These physical initial conditions were not known until now. Their
discovery calls for reconsideration of many well studied problems in
GR, including the gravitational collapse problem.

We present here an instant static solutions of the RCE with
singularities, which correspond to presence of arbitrary number of
massive matter points, both of discrete or of continuous
distribution. These solutions define the physically meaningful
initial conditions for EE, which describe the {\em real} matter,
made of massive point particles. They present a very special class
in the variety of all initial condition, which are admissible from
pure mathematical point of view.

To obtain the non-stationary gravitational field of moving matter
point sources, the instant solutions of RCE can be modified in a
manner, which is well known from relativistic electrodynamics. Of
course, the whole problem is a highly complicated and still not
solved. One can hope that in GR a procedure, which is analogous to
the introduction of Li\'eanard-Wiechert potentials may take place.
Here we shall stress that in electrodynamics for this purpose the
static Colomb potential is in use. We believe that in a similar way
our fundamental solutions may turn to be the key tool for the
correct treatment of the nonlinear GR dynamics of many-particle
systems, which is still an open problem.

As already mentioned, in GR, due to gravitational mass defect, we
have to distinguish two different masses of every body -- the
Keplerian gravitational mass $m$ and the proper bare mass $M>m$ of
the body \cite{books}. Under some additional natural assumptions our
new quasi-linear superposition principle for the fundamental
solution of EE yields a novel theory of the relativistic
gravitational mass defect of systems of discrete matter points and
composite bodies of continuous mass distribution. It is based on a
specific integral equation for the relativistic gravitational
potential, derived for the first time here. It turns out that the GR
mass defect is governed completely by the RCE, with time $t$,
playing the role of an auxiliary parameter in its solutions. Thus,
our quasi-linear superposition principle for the fundamental
solution of EE has a basic impact on the relativistic theory of
gravitational defect of mass. We give here for the first time a
number of solutions of the integral equation for the relativistic
mass defect. These describe some basic physical problems:
two-point-particle problem, some special cases of three and four
point particle problems, the general properties of the
$N$-point-particle problem, as well as some basic examples of
continuous mass distribution: homogeneous massive circles, spheres
and balls.

Further important physical consequences, which can be derived
using the new fundamental solutions of EE and the corresponding
nonlinear superposition principle for them, will be considered
elsewhere.

\section{The Mathematical Problem of Single Point Particle in GR}

\subsection{The Total Action and Introduction of Coordinates}

Let us suppose that in the whole universe there exist only a single
{\em massive} point particle of bare mechanical mass $M$, and that
it creates its own gravitational field according to the laws of GR.
This problem is described by the total action ${\cal
A}_{tot}\!=\!{\cal A}_{GR}+{\cal A}_{M}$. The first term describes
the action of the gravitational field, created by the single
particle. The second term adds to the total action the pure matter
(mechanical) action of the massive particle. Thus in GR the total
action acquires the well known explicit form:
\ben {\cal A}_{tot}\!=\!\!-{1\over{16\pi}}\int\!d^4 x\sqrt{|g|}R -
M\!\!\int\!ds. \la{total_action} \een

In the rest frame of the point particle both the total action
(\ref{total_action}) and the formed by this particle GR space-time
manifold $\mathbb{M}^{(1,3)}\{g_{\mu\nu}\}$ have an obvious group of
symmetry $SO(3)\times T_t(1)$. As a result, the problem can be
reduced not only on the orbits of the group $SO(3)$, i.e. on the 2D
quotient space $\mathbb{M}^{(1,1)}=\mathbb{M}^{(1,3)}/SO(3)$, with
natural global coordinates $t$ and $r$, but even on the orbits of
the whole group $SO(3)\times T_t(1)$, i.e. on the 1D quotient space
$\mathbb{M}^{(1)}=\mathbb{M}^{(1,3)}/\big( SO(3)\times T_t(1)\big)$,
with some natural radial coordinate $r$.

To be able to use some coordinates $x=\{x^\mu\}$ in the Riemannian
space-time $\mathbb{M}^{(1,3)}\{g_{\mu\nu}\}$ of the point particle
problem, one actually presupposes to have a flat Minkowskian
space-time $\mathbb{E}^{(1,3)}\{\eta_{\mu\nu}\}$, endowed with the
same coordinates. For example, one assumes to borrow  the Cartesian
coordinates: $\{t,{\bf r}\}$, or the spherical ones:
$\{t,r,\theta,\phi\}$ from the flat Minkowskian space-time for the
use in the Riemannian space-time. Thus we have at our disposal
simultaneously a flat metric $\eta_{\mu\nu}(x)$, and a Riemannian
metric $g_{\mu\nu}(x)$, expressed in the same coordinates.

One of the basic results of present article is that the auxiliary
flat space-time $\mathbb{E}^{(1,3)}\{\eta_{\mu\nu}\}$ plays much
more profound role in the problem at hand, than the usually expected
formal one. In particular, it turns out that the {\em real}
geometrical points of the two space-times:
$\mathbb{E}^{(1,3)}\{\eta_{\mu\nu}\}/{\cal W}_{{\bf 0}}$, with the
world line ${\cal W}_{{\bf 0}}$ of one point (the origin $r=0$)
removed, and $\mathbb{M}^{(1,3)}\{g_{\mu\nu}\}$ -- the space-time of
the GR massive point particle problem, are in one-to-one
correspondence. In particular, the last has the same topology as the
first one. Note that the Galilean space-time of a single point
particle problem in Newtonian gravity has precisely the same
topology as the one of the manifold
$\mathbb{E}^{(1,3)}\{\eta_{\mu\nu}\}/{\cal W}_{{\bf 0}}$. Thus, in
the real domain of variables, the space-time in the Newton gravity
and in the GR, formed by a single massive point particle, have the
same topology. This observation makes it clear that even because of
pure topological reasons the black hole solutions of EE are not
compatible with the matter point sources of gravity, sice they have
a different topology.

In its proper frame the single massive point particle, placed at the
origin of the standard spherical coordinate system in the 3D
Riemannian space
$\mathbb{M}^{(3)}\{g_{ij}\}\subset\mathbb{M}^{(1,3)}\{g_{\mu\nu}\}$
yields the familiar static metric (\ref{4Dinterval}) with three
unknown functions $g_{tt}(r)\geq 0,\,g_{rr}(r)\leq 0$, and
$\rho(r)\geq 0$ of the radial variable $r\geq 0$ \cite{books}. The
variable $r$ is not defined by the $SO(3)$ symmetry of the problem,
nor by its global-time translation invariance with respect to the
group $T_t(1)$. From geometrical point of view the choice of the
function $\rho(r)$ fixes the imbedding of the quotient space
$\mathbb{M}^{(1)}=\mathbb{M}^{(3)}/SO(3)$ into the 3D space
$\mathbb{M}^{(3)}$. We assume that by definition the value $r=0$ of
the radial variable $r$ corresponds to the center of spherical
symmetry, $C$. There the massive matter point is placed. We also
accept other assumptions about the mathematical properties of the
radial variable $r$, listed in the Subsection 1.2.2 of the
Introduction.

\subsection{On the Role of the Gauge Fixing in the Massive Point Problem}

General relativity is a gauge theory. The fixing of the gauge in
GR is described by a proper choice of the quantities
$$\bar\Gamma_\mu\!=\!-{{1}\over{\sqrt{|g|}}}g_{\mu\nu}
\partial_\lambda\left(\sqrt{|g|}g^{\lambda\nu}\right)$$
in the 4D d'Alembert operator
$g^{\mu\nu}\nabla_\mu\nabla_\nu=g^{\mu\nu}
\left(\partial_\mu\partial_\nu-\bar\Gamma_\mu\partial_\nu\right)$
\cite{books}, and actually is a fixing of the coordinates. In our
problem the choice of spherical coordinates and static metric
{dictates} the form of three of the quantities $\bar\Gamma_\mu$:
$\bar\Gamma_t\!=\!0,\,\,
\bar\Gamma_\theta\!=-\!\cot\theta,\,\,\bar\Gamma_\phi\!=\!0$, but
the function $\rho(r)$ and, equivalently, the form of the quantity
$$\bar\Gamma_r\!=\left(\!
\ln\left({\sqrt{-g_{rr}}\over{\sqrt{g_{tt}}\,\rho^2}}\right)\!\right)^\prime$$
{are still not fixed.} Here and further on, the prime denotes
differentiation with respect to the variable $r$. We refer to the
freedom of choice of the function $\rho(r)$ as {\em a rho-gauge
freedom}  in a broad sense, and to the choice of the $\rho(r)$
function  as {\em a rho-gauge fixing}.

At first glance the function $\rho(r)$ may be chosen in quite
arbitrary way, thus fixing the remaining (radial) gauge freedom of
the problem -- the only one, which is not fixed by symmetry
reasons. We show that choosing a definite class of functions
$\rho(r)$ one can solve correctly the EE (\ref{EE}) with
stress-energy tensor
\ben T^\nu_\mu=M\,\delta_{g}^{(3)}({\bf
r})\,\delta^\nu_0\,\delta^0_\mu= M{\delta^{(3)}({\bf
r})\over{\sqrt{|{}^3\!g({\bf r})|}}}\,\delta^\nu_0\,\delta^0_\mu,
\hskip .5truecm \la{Tmunu}\een
which describes a massive point source with bare mass $M$ at rest in
stationary and static coordinates. It may seem strange that for
solving this problem, one needs to fix the class of coordinates by a
proper choice of the radial gauge. As we shall see, the choice of
the admissible class of radial coordinates $r$ is a consequence of
the boundary conditions. In the problem at hand these conditions are
masked in 3D Dirac $\delta$-function in (\ref{Tmunu}). It describes
in a formal mathematical way the properties of source of gravity
{\em and} its boundary.

The following comments throw an additional light on this delicate
issue:

1. The strong believe in the independence of the GR results on the
choice of coordinates $x$ in the space-time
$\mathbb{M}^{(1,3)}\{g_{\mu\nu}(x)\}$ predisposes us to a somewhat
light-head attitude towards the choice of the coordinates for a
given specific problem. Indeed, it is obvious that physical results
of any theory must not depend on the choice of the variables and, in
particular, these results must be invariant under any admissible
changes of the coordinates. This requirement is a basic principle in
GR. It is fulfilled in any {\em already fixed} mathematical problem.

2. Nevertheless, the change of the {\em interpretation} of the
variables may change the formulation of the very mathematical
problem and thus, the physical results. This can happen, because we
are using the variables according to their meaning. For example, if
we are considering the luminosity distance $\rho$ as a radial
variable of the problem, it seems natural to put the point source at
the point $\rho=0$. In general, we may obtain a physically different
problem, if we are considering another variable $r$ as a radial one.
In this case we shall place the source at a different geometrical
point $r=0$, which now seems to be the natural position for the
center $C$, but does not coincide with the previous one -- $\rho=0$.
Imposing the same physical requirements, i.e. the same boundary
conditions {\em at different} places in the space, we obviously will
obtain different physical problems and results. Of course, as in any
gauge theory, in GR there exist a classes of physically equivalent
gauges. All gauges (coordinates) in such class yield the same
physical results. The real problem is how to find the correct class
of the gauges, proper for the given physical configuration.

3. The relation between the two geometrical "points": $\rho=0$ and
$r=0$, and between the corresponding physical models of a point
particle, strongly depends on the choice of the class of functions
$\rho(r)$, i.e. on the class of the radial gauges. Thus, applying
the same physical requirements in essentially different "natural"
variables, we arrive at different physical models, because we are
solving EE under different boundary conditions, coded in
corresponding 3D Dirac $\delta$-functions in (\ref{Tmunu}). One has
to find a theoretical or an experimental reasons to resolve this
essential ambiguity.

4. The choice of the radial coordinate in the single point particle
problem in GR needs a careful analysis. It is essential for the
description of the {\em very} source of gravitational field, not for
the description of the field in surrounding this source vacuum
domain. A well known mathematical fact is that in the vicinity of a
definite singular point of a mathematical functions one must use a
definite special type of coordinates for adequate description of the
character of the singularity, i.e., one is to solve the
corresponding uniformization problem.

5. The solutions of EE in {\em essentially} different coordinates
have different singularities somewhere {\em in the whole} complex
domain of the corresponding  variables. The essentially different
coordinates may be equivalent only locally -- in the spirit of the
widely used theory of smooth manifolds. One ought to make a
reservation, speaking about "essentially different coordinates",
because there exist a coordinate changes, which alter {\em only} the
place of the singularities of the solutions of EE in the complex
domain of the variables, without varying the character and the
number of these singularities. Such changes are precisely the linear
ones and the fractional-linear ones. All other, more general
coordinate changes, do not possess such property and yield
essentially different coordinates in the whole complex domain.

6. In our particular problem, according to Birkhoff theorem, the
spherically symmetric solution with given Keplerian mass $m$ is
unique {\em in the vacuum} domain. The coordinates, which are
essentially different somewhere else, may be locally equivalent in
the vacuum domain. As a result, all {\em local} GR effects, like
gravitational redshift, perihelion shift, deflection of light rays,
time-delay of signals, etc., will have their standard {\em exact}
values in static spherically symmetric gravitational field with
given Keplerian mass $m$. These physical values do not depend on the
admissible coordinate form of the solution.

We will use this local gauge freedom in description of the
gravitational field {\em outside} the source to reach an adequate
mathematical modelling of the very point source.

\subsection{The Gravitational Field Equations and Their Solution}

\subsubsection{The Vacuum Solution in an Arbitrary Radial Gauge}

The EE (\ref{EE}) for our problem with metric (\ref{4Dinterval})
can be easily derived from the following form of the nonzero
components of Einstein tensor:
\begin{subequations}\label{Gmunu_g:abc}
\ben
G^t_t\!=\!{1\over{-g_{rr}}}\left(\!-2\left(\!{{\rho^\prime}\over{\rho}}\!\right)^{\!\prime}
\!-\!3\left(\!{{\rho^\prime}\over{\rho}}\!\right)^{\!2}\!+
\!2{{\rho^\prime}\over{\rho}}{{\sqrt{-g_{rr}}^{\,\prime}}\over{\sqrt{-g_{rr}}}}
\right)\!+\!{{1}\over{\rho^2}},\hskip .15truecm  \la{Gmunu_g:a} \\
G^r_r\!=\!{1\over{-g_{rr}}}\left(\!-\!\left(\!{{\rho^\prime}\over{\rho}}\!\right)^{\!2}\!+
\!2{{\rho^\prime}\over{\rho}}{{\sqrt{g_{tt}}^{\,\prime}}\over{\sqrt{g_{tt}}}}
\right)\!+\!{{1}\over{\rho^2}},\hskip 2.4truecm  \la{Gmunu_g:b}\\
G^\theta_\theta\!=\! G^\phi_\phi\!=\!{1\over{-g_{rr}}}
\Bigg(\!\!-\!\left(\!{{\rho^\prime}\over{\rho}}\!\right)^{\!\prime}\!
\!-\!\left(\!{{\rho^\prime}\over{\rho}}\!\right)^{\!2}\!
\!-\!{{\rho^\prime}\over{\rho}}{{\sqrt{g_{tt}}^{\,\prime}}\over{\sqrt{g_{tt}}}}
\!+\hskip 1.45truecm \la{Gmunu_g:c}
\\
\!{{\rho^\prime}\over{\rho}}{{\sqrt{-g_{rr}}^{\,\prime}}\over{\sqrt{-g_{rr}}}}\!-\!
\left(\!{{\sqrt{g_{tt}}^{\,\prime}}\over{\sqrt{g_{tt}}}}\!\right)^{\!\prime}\!
\!-\!\left(\!{{\sqrt{g_{tt}}^{\,\prime}}\over{\sqrt{g_{tt}}}}\!\right)^{\!2}\!
\!+\!{{\sqrt{g_{tt}}^{\,\prime}}\over{\sqrt{g_{tt}}}}
{{\sqrt{-g_{rr}}^{\,\prime}}\over{\sqrt{-g_{rr}}}}
\Bigg),\nonumber\een
\end{subequations}
using, in addition, the corresponding components of the
energy-momentum tensor (\ref{Tmunu}) of a single matter point.

In particular, solving the EE for the static, spherically
symmetric case {\em in vacuum}, one easily obtains the following
most general solution:
\ben g_{tt}(r)=1-{{2m}\over\rho(r)}>0,\,
g_{rr}(r)=-\left(\rho^\prime(r)\right)^2/g_{tt}(r)<0,\nonumber\\
\rho(r)\,\hbox{\,--\, an arbitrary ${\cal C}^{1}$ function}.\hskip
3.7truecm \la{solution_geenral}\een
It was derived for the first time already in the articles
\cite{ComJann}.

If one uses the Hilbert gauge $\rho_H(r)=r$ in the EE with
$\delta({\bf r})$ term in the rhs, one easily reaches a
contradiction \cite{NP,PF}. Hence, now the question is as to how
to choose the radial-gauge-fixing-function $\rho(r)$, to be able
to comply with the specific boundary conditions at $r=0$, coded in
the $\delta({\bf r})$ term in the rhs of EE. In other words, we
have to find a radial gauge, which makes the boundary problem for
EE consistent with the presence of matter point source of gravity.

\subsubsection{Normal Coordinates for Static Spherically Symmetric
Gravitational Field}

The expressions (\ref{Gmunu_g:abc}) demonstrate a very important
feature of EE: In spite of their nonlinearity, which may yield
doubts in the applicability of the theory of mathematical
distribution, EE are quasi-linear differential equations.  After
all, the higher (second) order derivatives of the unknown functions
enter these equations linearly. This makes possible the usage of
mathematical distributions \cite{Gelfand} in the GR massive point
particle problem in some {\em specific} coordinates and the usage of
the Colombeau's theory of the generalized functions
\cite{Colombeau_GR}, hoppefully in all admissible coordinates.

The fundamental solutions of EE were found for the first time in
the articles \cite{PF}, introducing (in a slightly different
notations) proper normal field variables $\varphi(r)$,
$\varphi_2(r)$ and $\bar\varphi(r)$ according to the formulas
\ben g_{tt}=\exp(2\varphi),\,\,\,
\rho=\bar\rho\exp(-\varphi+\varphi_2),\,\,\,
g_{rr}=-\exp(-2\varphi+4\varphi_2-2\bar\varphi).\la{norm_var}\een
Here $\bar\rho=const>0$ defines the scale of the luminosity
variable.

In the present article we develop a more general approach to the
fundamental solutions of EE, which allows consideration of
arbitrary number of massive point sources of gravity. We shall see
that an essential ingredient of this approach remains the Fock
conformal transformation of the space $\mathbb{M}^{(3)}$, see Fock
in \cite{ books}. It arises naturally when one puts the
gravitational action of the static spherically symmetric problem
to a canonical form \cite{PF}. Indeed, after reduction of the
Hilbert-Einstein action ${\cal A}_{GR}$ on the orbits of the group
$SO(3)\times T_t(1)$, we arrive at one dimensional variational
problem with "Lagrangian"
\ben {\cal L}={\frac {1} {2}}\!\left({
{2\rho\rho^\prime\left(\sqrt{g_{tt}}\right)^\prime\!+\!
\left(\rho^\prime\right)^2\sqrt{g_{tt}} } \over {\sqrt{\!-g_{rr}}}
}\!+\!\sqrt{g_{tt}}\sqrt{\!-g_{rr}}\right).\la{LGR}\een
The corresponding Euler-Lagrange equations for pure gravitational
field in vacuum read:
\begin{subequations}\label{EL:abc}
\ben
\left({{2\rho\rho^\prime}\over{\sqrt{\!-g_{rr}}}}\right)^\prime
-{{{\rho^\prime}^2}\over{\sqrt{\!-g_{rr}}}}-\sqrt{\!-g_{rr}}=0,
\la{EL:a}\\
\left({{\left(\rho\sqrt{\!g_{tt}}\right)^\prime}\over{\sqrt{\!-g_{rr}}}}
\right)^\prime
-{{\rho^\prime\left(\sqrt{\!g_{tt}}\right)^\prime}\over{\sqrt{\!-g_{rr}}}}=0,
\la{EL:b}\\
{{2\rho\rho^\prime\left(\sqrt{g_{tt}}\right)^\prime\!+\!
\left(\rho^\prime\right)^2\sqrt{g_{tt}} } \over {\sqrt{\!-g_{rr}}}
}\!-\!\sqrt{g_{tt}}\sqrt{\!-g_{rr}}\stackrel{w}{=}0, \la{EL:c}\een
\end{subequations}
where the symbol "\,$\stackrel{w}{=}$\," denotes a weak equality
in the sense of theory of constrained dynamical systems. As a
result of the rho-gauge freedom the field variable
$\sqrt{\!-g_{rr}}$ is not a true dynamical variable but rather
plays the role of a (specific nonlinear) Lagrange multiplier,
which is needed in a description of constrained dynamics. Its
derivative with respect to the radial variable $r$ does not enter
the Lagrangian (\ref{LGR}). An advantage of such derivation of
field equation is that it makes transparent this fact. Of course,
the equations (\ref{EL:abc}) are completely equivalent to the
vacuum EE, considered in the previous subsection. As a result
equations (\ref{EL:abc}) are solved by the functions
(\ref{solution_geenral}).

Let us consider the formal 2D  space $\mathbb{M}^{(2)}$ of the
field variables  $\sqrt{g_{tt}}$ and $\rho$, endowed with the
quadratic metric form $ {2\rho\over \sqrt{\!-g_{rr}(\rho)} }
\,d\rho \,d\!\left(\sqrt{g_{tt}}\right)+{\sqrt{g_{tt}} \over
\sqrt{\!-g_{rr}(\rho)}} d\rho^2$. It is easy to check that its
Riemannian curvature tensor is zero. Hence, in this space one can
introduce a normal field variables, transforming its 2D metric
into canonical form. The above change of variables yields the
corresponding diagonal form of the Lagrangian:
\ben {\cal L}\!=\!\!{1\over{2}}\!
\Bigl(e^{\bar\varphi}\left(\!-(\bar\rho\varphi^\prime)^2\!+
\!(\bar\rho\varphi_2^\prime)^2\right)
\!+\!e^{\!-\bar\varphi}e^{2\varphi_2}\Bigr). \la{L2}\een
Hence, the new field variables play the role of {\em a normal}
fields' variables for the problem at hand. In these variables the
metric acquires the form
 \ben ds^2=
\!e^{2\varphi}dt^2\!\!-\!e^{\!-2\varphi\!+4\varphi_2-2\bar\varphi}dr^2\!\!-\!
\bar\rho^2e^{\!-2\varphi\!+2\varphi_2}(d\theta^2\!\!+
\!\sin^2\!\theta
d\phi^2)=\!e^{2\varphi}dt^2\!\!-\!e^{\!-2\varphi}dl_{\!{}_F}^{\,2}
\la{nc} \een where $\varphi(r)$, $\varphi_2(r)$ and $\bar\varphi(r)$
are still unknown functions of the variable $r$.

Obviously, the variable $\varphi$ describes the Fock conformal
transformation to the 3D space with infinitesimal distance
$dl_{\!{}_F}$. The variable $\bar\varphi$ is not a dynamical one and
fixes the radial gauge. We define a basic radial gauge (BRG) via the
relation $\bar\varphi_{BRG}(r)\equiv 0$. In BRG the coefficients of
the diagonal kinetic term in (\ref{L2}) are constant.

\subsubsection{Solution of Einstein Equations for Single Massive Point
Source}

{\bf a) Distributional form of EE for Point Particle.}

\vskip .2truecm

\noindent Let us consider EE (\ref{EE}), rewritten in the form
\ben R^\mu_\nu-8\pi\left(T^\mu_\nu-{1\over 2}
T\delta^\mu_\nu\right)=0.\la{EER}\een
Since the energy-momentum tensor (\ref{Tmunu}) of the problem is a
distribution: $T^\mu_\nu(x)\in {\cal
D}^\prime\mathbb{M}^{(1,3)}\{g_{\mu\nu}(x)\}$, the correct
mathematical treatment requires to consider tensor-valued test
functions $\Psi_\mu^\nu(x)\in {\cal
D}\mathbb{M}^{(1,3)}\{g_{\mu\nu}(x)\}$ \cite{TaubRaju} and to
rewrite equations (\ref{EER}) in the form
\ben \int_{\mathbb{M}^{(1,3)}}d^4 x\sqrt{|{}^4 g(x)|}
\left(R^\mu_\nu-8\pi\left(T^\mu_\nu-{1\over 2}
T\delta^\mu_\nu\right)\right)\Psi^\nu_\mu(x)=0. \la{EERD}\een
For a static problem the expression in the lhs in equation
(\ref{EER}) does not depend on the variable $x^0$ and we can use
test functions of the form $\Psi^\nu_\mu(x)=\psi^\nu_\mu({\bf r
})\chi^\nu_\mu(x^0)$, where
$\int_{\!-\infty}^{\,\,\infty}dx^0\chi^\nu_\mu(x^0)=1$ and
$\psi^\nu_\mu({\bf r}) \in {\cal D}\mathbb{M}^{(3)}\{g_{ij}({\bf
r})\}$. Then equation (\ref{EERD}) reduces to
\ben \int_{\mathbb{M}^{(3)}}d^3 {\bf r}\sqrt{g_{tt}({\bf r})\,|{}^3
g({\bf r})|} \left(R^\mu_\nu-8\pi\left(T^\mu_\nu-{1\over 2}
T\delta^\mu_\nu\right)\right)\psi^\nu_\mu({\bf r})=0. \la{EERD3}\een
These equations must be fulfilled for any test unctions
$\psi^\nu_\mu({\bf r})$.

\vskip .3truecm

{\bf b) Solution of the relativistic constraint equation (RCE).}

\vskip .2truecm

\noindent Taking into account:

i) the only nonzero component of the energy momentum tensor
(\ref{Tmunu}) $T_0^0=M\delta^{(3)}_g({\bf r})$;

ii) the expression
$R^0_0={1\over\sqrt{g_{tt}}}\Delta_g\left(\sqrt{g_{tt}}\right)$,
which is valid in the static case, where
$\Delta_g={1\over\sqrt{|{}^3g|}}\partial_i\left(\sqrt{|{}^3g|}g^{ij}\partial_j\right)$
is the Laplacean in $\mathbb{M}^{(3)}\{g_{ij}({\bf r})\}$; and

iii) using the specific test functions of the type
$\psi^\nu_\mu({\bf r})=\psi({\bf r})\,\delta^\nu_0\, \delta^0_\mu$;

\noindent we obtain the RCE on the initial conditions in the form
\ben \int_{\mathbb{M}^{(3)}}d^3 {\bf r}\sqrt{|{}^3 g({\bf
r})|}\Delta_g\left(\sqrt{g_{tt}}\right) \psi({\bf r})=4\pi
m\psi(0). \la{icc}\een
Here emerges a new constant
\ben m:=M\sqrt{g_{tt}(0)}. \la{m}\een

In spherical coordinates one easily finds $\sqrt{|{}^3 g({\bf
r})|}=\sqrt{-g_{rr}({\bf r})}\rho^2(r)/r^2$. Then using the normal
field's variables (\ref{norm_var}) and 3D Euclidean space
notations, we reach the final form of the RCE:
\ben \bar\rho^{2}\!\int_{\mathbb{E}^{(3)}}\! d^3{\bf r}\,
\nabla\!\cdot\!\left( \varphi^\prime e^{\bar\varphi}
\nabla\left({{-1}\over r}\right)\right) \psi({\bf r})= 4\pi
m\psi(0).\la{iccE}\een
Its solution determines the dependence of the function
$\varphi(r)$ on the radial gauge function $\bar\varphi(r)$ in the
form:
\ben \varphi(r)={m\over{\bar\rho^2}}\int_{r_{\infty}}^r dr
e^{-\bar\varphi(r)}.\la{solution}\een
The value $r_\infty>0$ of the radial variable, used in formula
(\ref{solution}), defines the place, where $\varphi(r_\infty)=0$,
i.e., where $g_{tt}(r_\infty)=1$.  This value $r_\infty$ obviously
may depend on the choice of the gauge function $\bar\varphi(r)$.
The value of the $\varphi(r)$ at the place of the point source is
\ben \varphi(0)=-{m\over{\bar\rho^2}}\int^{r_{\infty}}_0 dr
e^{-\bar\varphi(r)}<0.\la{phi0}\een

Now we see that the solution of RCE translates the differential
3-form $\omega^3_g:=d^3 {\bf r}\,\sqrt{|{}^3 g({\bf
r})|}\Delta_g\left(\sqrt{g_{tt}}\right)$ on
$\mathbb{M}^{(3)}\{g_{ij}({\bf r})\}$ to the distribution-valued
differential 3-form $\omega^2_\delta:=d^3{\bf
r}\,\Delta\left({{-m}\over r}\right)=d^3{\bf r}\, m\delta({\bf r})$,
defined on the Euclidean space $\mathbb{E}^{(3)}\{\delta_{ij}\}$,
i.e., on the solution of RCE we have the relation
\ben
\omega^3_g=\omega^3_\delta.\la{omega}\een
One can consider this relation (\ref{omega}) as a new form of the
RCE.

The correspondence between the spaces
$\mathbb{M}^{(3)}\{g_{ij}({\bf r})\}$ and
$\mathbb{E}^{(3)}\{\delta_{ij}\}$ was stressed in Section 2.1. The
extension of this correspondence, obtained here, is the
geometrical basis for application of the mathematical theory of
distributions in the massive point particle problems in GR.

\vskip .3truecm

{\bf c) Solution of the Rest of  EE.}

\vskip .2truecm

Since the other components of the energy-momentum tensor of point
particle are zero in its proper frame, one can use for them any of
the forms (\ref{EE}), \ref{EER}), or (\ref{EERD}) of EE, and we
arrive at the following ordinary differential equations for normal
field variables:
\begin{subequations}\label{Eqphi:ab}
\ben \varphi^{\prime\prime}_2+\bar\varphi^\prime\varphi^\prime_2=
{1\over{\bar\rho^2}}e^{2(\varphi_2-\bar\varphi)},\label{Eqphi:a}\\
\left(\varphi^\prime\right)^2-\left(\varphi^\prime_2\right)^2 +
{1\over{\bar\rho^2}}e^{2(\varphi_2-\bar\varphi)}\stackrel{w}{=}0\label{Eqphi:b}
\la{phi2}\een
\end{subequations}
Using the relation (\ref{solution}), we can exclude the gauge
function $\bar\varphi$ from this system, obtaining its
radial-gauge-invariant form:
\begin{subequations}\label{Eqphi_inv:ab}
\ben {{d^2\varphi_2}\over{d\varphi^2}}=
{{\bar\rho^2}\over{m^2}}\,e^{2\varphi_2},\label{Eqphi_inv:a}\\
1- \left({{d\varphi_2}\over{d\varphi}}\right)^2 +
{{\bar\rho^2}\over{m^2}}\,e^{2\varphi_2}\stackrel{w}{=}0\label{Eqphi_inv:b}
\la{phi2}\een
\end{subequations}
Note that meanwhile we have excluded the radial-gauge-dependent
variable $r$ replacing it with the radial-gauge-independent one --
$\varphi$, which now plays the role of the independent "radial"
variable.

The first equation (\ref{Eqphi_inv:a}) can be integrate immediately
in quadratures. The second equation (\ref{Eqphi_inv:b}) imposes a
constraint on the two integration constants in the general solution
of the first one. Thus we remain with only one integration constant
$\varphi_\infty$ in the solution of the system (\ref{Eqphi_inv:ab}):
\ben \varphi_2(\varphi)=\ln\left({{m/\bar\rho}\over
{|\sinh\left(\varphi-\varphi_\infty\right)|}}\right), \,\,\,\,\,
\rho(\varphi)={{m\exp(-\varphi)}\over{|\sinh\left(\varphi-\varphi_\infty\right)|}}.
\la{sol_phi2_rho}\een
\vskip .3truecm

{\bf d) Fixing the emerging constants in the general solution of
the problem.}

\vskip .2truecm

\noindent The second expression in equations (\ref{sol_phi2_rho})
is derived using formulas (\ref{norm_var}). It shows that
$\rho(\varphi_\infty)=\infty$. Hence, the value $\varphi_\infty$
corresponds to the physical infinity, where the space-time is
asymptotically flat and we must have $g_{tt}=1$. Thus we see that
the value $\varphi_\infty$ must be reached for the value of the
radial variable $r_\infty$, i.e. the relation
$\varphi_\infty=\varphi(r_\infty)=0$ must take place. For the
value of the radial variable $r_\infty$ the space-time
$\mathbb{M}^{(1,3)}\{g_{\mu\nu}(x)\}$ is asymptotically flat. This
value corresponds to the physical infinity. Hence, under our
conventions, described in the Introduction, the physically
admissible interval of the values of the radial variable is
$r\in[0,r_\infty]$.

As a result we remain with only two arbitrary constants $m$ and
$r_\infty$ in the solution of the whole system of EE for the massive
point particle. The 4-D metric in normal field's variables acquires
the final gauge invariant form in which its coefficients are
functions only of $\varphi$ in the role of a radial variable:
\ben
ds^2=e^{2\varphi}dt^2-e^{-2\varphi}\,m^2\!\left({{d\varphi^2}\over{\sinh(\varphi)^4}}+
{{d\theta^2+\sin(\theta)^2d\phi^2}\over{\sinh(\varphi)^2}}\right).
\la{solution_metric}\een
It's remarkable that in the metric (\ref{solution_metric}) appears
only the integration constant $m$. As a result only the value of
this constant will influence the local dynamics of any {\em test}
particles and fields, which probe the metric in the space-time of
single massive point source. This important conclusion is
independent of the choice of radial variable $r$, i.e. it is gauge
invariant, as well as the whole equation (\ref{solution_metric}).

According to the definition (\ref{m}) of the constant $m$ and the
formulas (\ref{norm_var}) we have $\varphi(0)=\ln(m/M)$. The
relation (\ref{phi0}) shows that $\varphi(0)<0$. Thus we obtain,
that in the GR massive point particle problem the variable $\varphi$
varies in the interval $\varphi\in [\ln(m/M),0]$. After all, the
bare mechanical mass $M$ shows up in GR massive point particle
solution, defying the interval of the physical values of the
variable $\varphi$.

It is convenient to introduce the mass ratio
\ben \varrho=m/M\in (0,1).\la{varrho}\een
Then in the problem at hand the basic quantity
$\sqrt{g_{tt}}=\exp(\varphi)$ varies in the physical interval
$\sqrt{g_{tt}}\in [\varrho,1]$.

The final formula
\ben
\rho(\varphi)={{2m}\over{1-\exp(2\varphi)}}\geq{{2m}\over{1-\varrho^2}}
> 2m\la{rho}\een
shows that the luminosity variable $\rho$ in the gravitational
field of massive point particle cannot take values, less than
${{\rho_G}\over{1-\varrho^2}}>\rho_G$, since $2m=\rho_G$, as we
shall see in the next subsection. This is in strong contrast to
the situation with Schwarzschild solution in Hilbert gauge
(\ref{Hilbert}) and in complete accord with Dirac's suggestion
\cite{Dirac}.

\vskip .3truecm

{\bf d) Fixing of the physical radial variable.}

\vskip .2truecm

The previous consideration gives a correct mathematical ground for
our conclusion about the topology of the space-time of single
massive point particle problem in GR, as described in Section 2.1.
Indeed, it is easy to obtain from EE (\ref{EER}) the 4D scalar
curvature of space-time with single massive point source:
\ben {}^4\!R=-8\pi M\delta_g^{(3)}({\bf r}), \la{4R_point}\een
and its 3D scalar curvature
\ben {}^3\!R=16\pi M\delta_g^{(3)}({\bf r}).\la{3R_point} \een

Clearly, the last equation is the concrete form of the RCE
(\ref{3R_eq}) in the point particle case. Hence, the 3D space
$\mathbb{M}^{(3)}\{g_{ij}({\bf r})\}$ has a strong singularity at
the geometrical point $r=0$, where the massive matter point is
placed, and the solution of EE cannot be extended behind this point,
both from physical and from mathematical reasons. The space-time of
the problem $\mathbb{M}^{(1,3)}\{g_{\mu\nu}(x)\}$ has a singular
line -- the world line ${\cal W}_0$ of the massive matter point.
This singular line must be removed from the manifold
$\mathbb{M}^{(1,3)}\{g_{\mu\nu}(x)\}$ and we remain with the
topology, described in Section 2.1.

The simple correspondence between the Riemannian space-time of
point particle and Minkowskian space-time gives a good reason to
adopt, as much as possible, the basic properties of the
Minkowskian radial variable $r$ for the Riemannian case, as
described in Section 1.2.2. Then, according to results in the
previous Section, we have to fix the gauge function
$\bar\varphi(r)$ in such a manner, that as a result of relation
(\ref{solution}) we will have the mapping:
\ben  [\ln(m/M),0]_\varphi \stackrel{\varphi(r)}{ \longrightarrow}
[0,\infty]_r \la{phi_r_map}\een

A very important additional requisite of the mapping
(\ref{phi_r_map}) is the requirement to preserve the number and
the character of the original singularities of the solution
(\ref{solution_metric}) in the whole compactified complex domain
of variable $\varphi$. The mapping (\ref{phi_r_map}) is allowed
only to change the positions of these singularities in the
compactified complex domain of variable $r$.

The only way to fulfill this requirement is to use a
fractional-linear function $\varphi(r)={{ar+b}\over{r+c}}$ with some
constant coefficients $a,b,c$, which are unambiguously fixed by the
mapping (\ref{phi_r_map}) in the form:
$\varphi(r)={b\over{r+b/\ln{\varrho}}}$. Taking into account that at
$r\to\infty$ the asymptotic of the function $\varphi(r)\sim b/r$
yields an asymptotic $g_{tt}\sim 1+2b/r$, one sees that the standard
comparison with the real observations imply $b=-m_{Kepler}$, where
$m_{Kepler}$ is the Keplerian mass of the particle, as observed by a
distant observer. Then the formula (\ref{rho}) gives
$\lim\limits_{r\to\infty}{{\rho(r)}\over r}=m/m_{Kepler}$ and the
Edington's coherent scale condition:
$\lim\limits_{r\to\infty}{{\rho(r)}\over r}=1$ (see Section 1.2.2)
fixes the value of our integration constant $m=m_{Kepler}$. Hence,
in this physical gauge the function
\ben \varphi(r)=-{{m}\over{|{\bf r}-{\bf r}_0|
+m/\ln{1\over\varrho}}}=-{{M\varrho}\over{|{\bf r}-{\bf
r}_0|+R}}=\varphi(r;M,R,\varrho) \la {phiGR}\een
presents a proper GR generalization of the Newton potential
(\ref{FSN}) $\varphi^{{}^{{}_{Newton}}}({\bf r})=-{{m}\over{|{\bf
r}-{\bf r}_0|}}$ of matter point with Keplerian mass $m$, placed
at the position ${\bf r}_0$, which describes the fundamental
solution in Newton theory of gravity\footnote{The form of the
Newton potential can be derived, following the same consideration
with only one difference:  the physical values of the Newton
potential cover the whole semi-constrained interval
$\varphi^{{}^{{}_{Newton}}}\in (-\infty,0]$. This interval has to
be mapped onto the interval $[0,\infty)\ni r$ by fractional-linear
function $\varphi^{{}^{{}_{Newton}}}(r)$ and this gives
$\varphi^{{}^{{}_{Newton}}}(r)=-{m\over r}$.}.

Now we see that the ratio $\varrho=m/M\in (0,1)$ describes the
relativistic gravitational defect of mass for massive point
particle. It was introduced for the first time in \cite{PF}, where
the solution of the problem was derived using a different
mathematical technic. As a final result we obtain a two parameter
family of solutions (\ref{phiGR}) to the massive point particle
problem in GR. This family can be parameterized by any two of the
three constants $m$, $M$, and $\varrho$.

In the formula (\ref{phiGR}) we use a short notation
$R=m\big/\ln{1\over\varrho}  = M\big/\left( {1\over
\varrho}\ln{1\over\varrho}\right)$ for the GR correction to the
Newton potential. Further on we shall refer to the correction $R$
as {\em "a relativistic shift"} in the Newtonian potential. The
potential $\varphi^{{}^{{}_{Newton}}}({\bf r})$ can be derived as
a limit $R\to 0$ of the relativistic one
(\ref{phiGR})\footnote{One easily obtains the following
instructive estimates: a) If $\varrho\lesssim 1/\sqrt{e}\approx
0.60653$, then $R\lesssim \rho_G$. b) If, according to Birkhoff's
theorem, one applies the formula (\ref{phiGR}) outside the
spherically symmetric body of finite radius $r_B$, the quantity
$R$ will not exceed the radius of the body $r_B$, when
$\varrho_B\lesssim exp(-\rho_G/r_B)$. This restriction is very
weak, since $\rho_G/r_B\lll 1$ for real bodies, and for any of
them $1-exp(-\rho_G/r_B)\lll 1$.}. Therefore we shall refer to the
very function $\varphi({\bf r})$ (\ref{phiGR}) as {\em "a
relativistic gravitational potential"}. From mathematical point of
view the relativistic gravitational potential defines the Fock
conformal mapping. It is clear that this potential plays a basic
physical role in the relativistic theory of gravity.

\subsubsection{Some Remarks on the Non-Relativistic Limit $c\to \infty$.}

One can expect that in the non-relativistic limit $c\to \infty$ our
GR  solution for single particle will reproduce the results of the
Newton theory. For study this limit it is necessary to restore the
physical units in corresponding formulae. Then we obtain for the
mass ratio:
\ben \varrho=e^{\varphi(0)/c^2}\la{varrho_c}, \een
and
\ben \varphi(r)=-{{G^{Newton}\, M\, e^{\varphi(0)/c^2}}\over{|{\bf
r}-{\bf r}_0| - G^{Newton}\, M\, e^{\varphi(0)/c^2}/\varphi(0)}} \la
{phiGR_c}\een
-- for the relativistic potential.

Unfortunately, at present we do not have a theory of the
relativistic collapse, which has to describe in detail the origin of
the relativistic gravitational defect of mass of a single point
particle and the value of the mass ratio $\varrho$. Here we are
considering $\varrho$ just as an additional free parameter of the
class of point particle solutions, studied in the present article.
If one considers, instead, as a free parameter $\varphi(0)$ and
assumes that it does not depend on the velocity of light, in the
limit $c\to \infty$ one obviously obtains from the relation
(\ref{varrho_c}) the limit:
\ben lim_{c\to \infty}\, \varrho=1. \la{varrho_c_to_infty} \een
This result sounds physically right. Indeed, one expects that in the
non-relativistic limit $c\to \infty$ the gravitational mass defect
will disappear and we will return back to the Newton theory of
gravity with $m=M$.

Despite of this physically reasonable result, the assumption that
$\varphi(0)$ does not depend on the velocity of light $c$ gives a
wrong limit
$ lim_{c\to \infty}\,\varphi(r)=-{G^{Newton}\, M}/\left({|{\bf
r}-{\bf r}_0| - G^{Newton}\, M/\varphi(0)}\right)$ in the formula
(\ref{phiGR_c}).

To obtain the physically right results in the both cases, one has to
assume that:

i) actually the quantity $\varphi(0)$ depends on the velocity of
light $c$ in some specific way, and

ii) the unknown at present function $\varphi(0,c)$ fulfills
simultaneously two additional conditions:
\ben lim_{c\to \infty}\,\,\varphi(0,c)=
\varphi^{Newton}(0)&=&-\infty,\nonumber\\
lim_{c\to \infty}\,\left(\varphi(0,c)/c^2\right)&=&0.
\la{varphi0_c_limits}\een

As a result of these conditions, which are obviously compatible, one
obtains both the relation (\ref{varrho_c_to_infty}) and the right
non-relativistic limit
\ben lim_{c\to \infty}\,\varphi(r)=-{{G^{Newton}\, M}\over{|{\bf
r}-{\bf r}_0|}}=\varphi^{Newton}(r).\la{{phiGR_c_to_infty}}\een

One can hope that the future theory of the relativistic
gravitational collapse, accompanied by a proper treatment of the
gravitational mass defect, or some other additional considerations,
will be able to derive the precise form of the function
$\varphi(0,c)$ and to confirm the physically natural relations
(\ref{varphi0_c_limits}). In the present article we will assume
these relations to be fulfilled.

\subsubsection{Some Remarks on the Properties of Static Fundamental Solutions}

{\bf 1. On the three dimensional form of the fundamental solutions
of EE.}

An unexpected and remarkable feature of the relativistic
gravitational potential (\ref{phiGR}) is that it has {\em a finite}
negative value $\varphi({\bf r}_0)=ln{\,\varrho}$ at the place of
the very point source ${\bf r}_0$. This unique property is in a
sharp contrast to the case of the Newton potential
$\varphi^{{}^{{}_{Newton}}}({\bf r})$ (\ref{FSN}), which diverges as
${-m/{|{\bf r}-{\bf r}_0|}}$, when ${\bf r}\to{\bf r}_0$. Thus we
see that in GR we have a self-regularizing mechanism for
gravitational interaction, based on the influence of matter on the
space-time curvature. In the article \cite{PFSD} we have shown that
the same phenomenon comes into being in GR electrostatic problem of
single massive point charge.

As a result of the relativistic self-regularization all components
of the metric tensor
$\sqrt{g_{tt}}=e^{\varphi},\,\,\,\sqrt{-g_{rr}}={{\varphi^2
e^{-\varphi}}\over{\sinh(\varphi)^2}},\,\,\,\rho={{m
e^{-\varphi}}\over{\sinh(-\varphi)}}$
in spherical coordinates are regular at the place of the point
source ${\bf r}_0$:
\ben \sqrt{g_{tt}({\bf r}_0)}=\varrho,\,\,\,\sqrt{-g_{rr}({\bf
r}_0)}={4\over\varrho}\left({{\varrho \ln\varrho}
\over{1-\varrho^2}}\right)^2,\,\,\,\rho({\bf
r}_0)={{2m}\over{1-\varrho^2}}.\la{metric_coeff_sph_0}\een

Now it becomes clear that the singular term $\delta^{(3)}({\bf r
}-{\bf r}_0)$ in the lhs of Eq. (\ref{iccE}) originates from the
singularity of the 3D Cartesian determinant:
\ben \sqrt{|{}^3g({\bf r})|}=\sqrt{-g_{rr}({\bf r})}\rho({\bf
r})^2/|{\bf r}-{\bf r}_0|^2\sim {1/{|{\bf r}-{\bf
r}_0|^2}},\,\,\,\hbox{when}\,\,\, {\bf r}\to{\bf
r}_0.\la{Cartesian_determinant}\een

The singularity of metric coefficients at the place of the point
source of gravity becomes transparent in Cartesian coordinates.
Indeed, one can write down the 3D distance in the tensorial form
$dl^2=-d{\bf r}\, \widehat{\,{}^3\!g({\bf r})}\, d{\bf r}$, using
the 3D Cartesian metric tensor
\ben -\widehat{\,{}^3\!g({\bf r})}={{\rho^2{(\bf r})}\over{{|\bf
r}-{\bf r}_0|^2}} \left(\mathbb{I}-{\bf e}_r\otimes{\bf
e}_r\right)-g_{rr}({\bf r})\,{\bf e}_r\otimes{\bf e}_r,
\,\,\,\hbox{where}\,\,\, {\bf e}_r:={{{\bf r}-{\bf
r}_0}\over{|{\bf r}-{\bf r}_0|}}. \la{3Dmetric}\een
The components of this tensor are obviously singular at the point
${\bf r}={\bf r}_0$. This is precisely because in the specific
geometry, defined by fundamental solutions of EE, we have
$\rho({\bf r}_0)>0$. The expression (\ref{Cartesian_determinant})
defines the square root of the determinant of tensor
(\ref{3Dmetric}).

\vskip .3truecm

{\bf 2. One-dimensional-like representation of the fundamental
solutions of EE.}

The Dirac $\delta$-function is a linear functional. Its
representation depends on the class of the test functions in use.
We can take advantage of spherical coordinates in description of
the test functions of the GR point particle problem. Starting with
Cartesian coordinate test functions $\psi({\bf r})\in {\cal
D}^\prime \{\mathbb{R}^{(3)}\}$, in spherical coordinates we
obtain a specific class of test functions
$\psi_{sph}(r,\theta,\phi):=\psi(r{\bf e}_r(\theta,\phi))\in {\cal
D}^\prime_{sph} \{\mathbb{R}^{(1)+}_r\times SO(3)\}$. These must
be distinguish form the arbitrary test functions
$\psi(r,\theta,\phi)=\psi(r,{\bf e}_r(\theta,\phi))\in {\cal
D}^\prime \{\mathbb{R}^{(1)+}_r\times SO(3)\}$ on the manifold
$\mathbb{M}^{(3)}=\mathbb{R}^{(1)+}_r\times SO(3)$. Now $r\in
\mathbb{R}^{(1)+}_r$ is considered as an independent variable, not
just as a short notation for $|{\bf r}|:=\sqrt{x^2+y^2+z^2}$. The
difference between the functions
$\psi_{sph}(r,\theta,\phi):=\psi(r{\bf e}_r(\theta,\phi))$ and
$\psi(r,\theta,\phi)=\psi(r,{\bf e}_r(\theta,\phi))$ is of
critical importance for our problem, since
$\psi_{sph}(r=0,\theta,\phi):=\psi({\bf 0})=const$ for any values
of the angle variables $\theta$ and $\phi$. The functions
$\psi(r,\theta,\phi)=\psi(r,{\bf e}_r(\theta,\phi))$ do not have
such property. Instead, the functions
$\psi(r=0,\theta,\phi)=\psi(0,{\bf e}_r(\theta,\phi))\neq const$
keep the dependence on the variables $\theta$ and $\phi$.

Let us use the class of test function ${\cal D}^\prime_{sph}
\{\mathbb{R}^{(1)+}_r\times SO(3)\}$. The standard restriction of
the Euclidean Laplacean  $\Delta \varphi(r)= {1\over
r}\,\partial^2_{r^2}\big(r\varphi(r)\big)$ on the functions, which
depend only on variable $r$, brings us to 1D formulation of the
problem. One can write down the solution of the Eq. (\ref{iccE})
with added point source of gravitational field, described by
function $\delta(r)$, in the following one-dimensional form  (see
for details \cite{PF}):
\ben \varphi(r)=\ln\varrho \left(1- {{r}\over{
r+R}}\,\Theta\!\left(\!{{r}\over{
r+R}}\!\right)\right).\la{Sol_1D}\een
Here we are using the Heaviside steep function $\Theta(r)$ with
regularization $\Theta(0)=1$.

The fundamental solutions of EE were found for first time in
\cite{PF} in this form. Its advantage is that it makes transparent
the jump in the derivatives of the metric coefficients in
spherical coordinates. This jump reproduces via the Einstein
tensor the $\delta$-function in the rhs of EE with point source.

In the case of 1D representation (\ref{Sol_1D}) the form of the
metric (\ref{solution_metric}) must be considered as valid only in
the vacuum domain, outside the point source. In this domain the
form (\ref{solution_metric}) does not make difference between 3D
and 1D representation of the fundamental solutions.

\subsubsection{On the Choice of Radial Gauge in the Single Particle Problem}

The above  consideration solves on a clear theoretical basis the
longstanding problem of the choice of radial variable $r$ for
point source of gravity in GR. The unambiguously obtained physical
radial variable $r$ is obviously a preferable one, both from
mathematical and from physical point of view.

The singularities of the metric coefficients in the whole
compactified complex plain $\widetilde{\mathbb{C}}_\varphi$ are
placed at the positions $\varphi_n=i\pi n,\,\,\,n\in \mathbb{Z}$.
The points of finite $n$ are poles and the infinite point
$|n|=\infty$ is an essentially singular one.

The singular points of the solution in the whole compactified
complex plain $\widetilde{\mathbb{C}}_r$ of the physical variable
$r$ are of two essentially different types:

1. The place of the point source of gravity at $r=0$ where the
curvature of space-time has a strong singularity, proportional to
$\delta^{(3)}({\bf r})$. This singularity is seen in the
differential 3-forms (\ref{omega}), too. Surprisingly, the
relativistic potential (\ref{phiGR}) and the metric coefficients,
when written in 3D form, are regular at this point.

2. In Einstein theory of gravity an unavoidable singular points of
the metric coefficient $g_{tt}(r)$ are the (complex) points
$r_n=-R+i {m\over {\pi n}},\,\,\,n\in \mathbb{Z}$. These singular
points are placed in the nonphysical domain of the physical variable
$r$. For finite $n\in \mathbb{Z}$ the singular points are poles of
the 3D metric coefficients in the expression
(\ref{solution_metric}). The sequence of these singular points has a
limiting point $r=-R$ for $|n|=\infty$. This is a real essentially
singular point, which is not an isolated one. In contrast, the real
singular point $r=-R$ is a simple pole of the relativistic potential
(\ref{phiGR}), which has no other singular points. This pole is
placed in {\em non-physical} domain $r<0$ and corresponds to the
real pole at $r=0$ of the gravitational potential in Newton theory
of gravity.

The multiplication of this single simple pole of the relativistic
potential $\varphi(r;M,R,\varrho)$ to an infinite series of
singularities of GR metric coefficients is produced by the specific
exponential mapping (\ref{solution_metric}) of
$\varphi(r;M,R,\varrho)$ onto these coefficients. This is a specific
feature of the relativistic description of gravitational field of
massive point particle and may be considered as a price, one has to
pay, for the self-regularization mechanism, discussed in the
previous subsection.

Under gage transformation to some other radial variable
$r_{other}$, related with the physical one $r$ by a coordinate
transformation $r_{other}=r_{other}(r)$ of general type, which is
not fractional-linear one, in the solution of the problem will
appear additional nonphysical singularities in the corresponding
compactified complex plain $\widetilde{\mathbb{C}}_{r_{other}}$.

From the relation (\ref{solution}) one easily obtains the physical
gauge function $\bar\varphi_{phys}(r)= 2\ln\left(
{{(r+R)}/{\bar\rho}} \right)$, which is compatible with boundary
conditions of the problem and with all additional requirements on
the physical radial variable $r$, as formulated in the
Introduction.

We shall call {\em a regular gauges} of the problem all gauges,
for which the integral (\ref{solution}) makes sense in the
physical interval $r\in [0,r_\infty]$. This is just the necessary
and sufficient condition for the radial gauge to be compatible
with the boundary conditions, coded in the Dirac $\delta$-function
in the RCE of the massive point particle problem.

This condition fixes a large class of admissible gauges for this
problem. One of them is BRG $\bar\varphi(r)\equiv 0$ in which the
relativistic potential has the form
$\varphi_{BRG}(r)=m(r-r_\infty)/\bar\rho^2,\,\,\,r\in [0,r_\infty]$.

Between the regular gauges for the one-particle problem in GR is the
gauge by Droste \cite{HDW}. It has a clear geometrical meaning,
since in this gauge the radial variable $r$ measures the radial {\em
geometrical distance} in the 3D Schwarzschild metric. This gauge
reproduces only a very special value of the mass defect ratio
$\varrho=(\sqrt{5}-1)/2\approx .6180$ \cite{PF}. Quite curiously,
under such geometrical choice of the radial gauge $\varrho$ equals
precisely the famous mathematical golden ratio.

All other known radial gauges, probed for spherically symmetrical
static solutions of EE and described in \cite{PF}, are not regular.
Therefore they cannot be used for solution of the point mass
problem. As we have seen in the previous subsections, only a certain
combination of gauge function $\bar\varphi(r)$ and corresponding
form of the relativistic potential $\varphi(r)$ can obey the
specific boundary conditions of this problem.

For example, in the most popular at present Hilbert gauge:
$\rho_{\!{}_H}(r)\equiv r$ the static spherically symmetric problem
has a relativistic potential
$\varphi_{\!{}_H}(\rho)=\ln\sqrt{1-\rho_G/\rho}$ and a radial gauge
function
$\bar\varphi_{\!{}_H}(\rho)=\ln\big(\rho(\rho-\rho_G)/(\bar\rho)^{\,2}\big)$.
Hence, for Hilbert gauge the integral (\ref{solution}) diverges
logarithmically: 1) when $\rho$ approaches the center $\rho=0$,
where the point source of gravity must be placed in this gauge, if
one insist on the point particle interpretation of the this form of
the Schwarzschild solution; 2) when $\rho$ approaches the event
horizon $\rho=\rho_G$. In addition, the value of this integral
becomes an imaginary number for $\rho<\rho_G$.

This means that Hilbert gauge is incompatible with the specific
boundary conditions for EE in presence of massive point particle.
Therefore one cannot use the Hilbert gauge to solve the point
particle problem in GR. This gauge yields the well known {\em
nonphysical} singularity at the point $\rho=0$, i.e. on the boundary
of the to-be-physical domain of the radial variable $\rho\in
[0,\infty)$. More over, the meaning of the variable $\rho$ radically
changes in the interval $[0,\rho_G]$. Here it plays the role of a
specific time variable and the point $\rho=0$ describes the future
infinity of the internal time $t_{in}=x-1/x\in (-\infty,\infty)$,
where $x\!=\!\rho\!+\!\rho_G\ln\left(|\rho/\rho_G\!-\!1|\right)$ is
the Regge-Wheeler "tortoise" coordinate in the interior of Hilbert
solution \cite{Fiziev2006}. It becomes clear that even if we will be
able to reproduce mathematically a term $\sim \delta(\rho)$ in the
rhs of EE (see the articles by P.~Parker, by H.~Belasin \&
H.~Nachbagauer, and by J.~M.~Heinzle \& R.~Steinbauer in
\cite{Colombeau_GR}), its interpretation as a source of the
gravitational field and the curvature of the Schwarzschild solution
is physically unacceptable. Such term may describe only a
$\delta$-shaped -- with respect to the time, "impulse" at the time
instant $\rho=0\,\, (\Rightarrow t_{in}=\infty)$ and has a complete
unclear physical meaning. In any case it is not able to describe the
usual physical 3D-space-point source of static gravitational field.

\subsection{The Total Energy of the Aggregate of Massive Point Source and its
Gravitational Field}

In the problem at hand we have an extreme example of an "island
universe``. In it a privileged reference system and a well defined
global time exist. It is well known that under these conditions
the energy of the gravitational field can be defined unambiguously
\cite{books}. Moreover, we can calculate the total energy of the
aggregate of a mechanical particle and its gravitational field in
a canonical way, considering the corresponding 1D variational
problem for total action (\ref{total_action}) in the spherically
symmetric static case \cite{PF}. The canonical procedure produces
a total Hamilton density ${\cal
H}_{tot}=\Sigma_{a=1,2;\mu=t,r}\,\pi_a^\mu\,\varphi_{a,\mu}-{\cal
L}_{tot}\!=\!{1\over{2}}\left(-\bar\rho^2{\varphi^\prime}^2
+\bar\rho^2{\varphi_2^\prime}^2-e^{2\varphi_2}\right)+M
e^{\varphi}\delta(r)$. Using the equations (\ref{Eqphi:ab}), one
immediately obtains for the total energy of the GR universe with
one point particle in it: \ben E_{tot}=\int_0^{{\infty}}{\cal
H}_{tot} dr=m=\varrho M< M\,.\la{E}\een

This result completely agrees with the strong equivalence
principle of GR. The energy of the {\em static} longitudinal
gravitational field, created by a point particle at rest is a
negative quantity:
$E_{GR}=E_{tot}-E_{\!{}_M}=m-M=-M(1-\varrho)<0$. Since both matter
point and its gravitational field have nonzero proper energies,
this result proves that the ratio $\varrho$ must belong to the
{\em open} interval $(0,1)$, see \cite{PF} for more details.

The above consideration gives a clear physical explanation of the
gravitational mass defect of a point particle.

\section{Quasi-Linear Superposition Principle for
Static Fundamental Solutions of Einstein Equations}

\subsection{Justification of the Quasi-Linear Superposition Principle
for the Relativistic Gravitational Potential $\varphi$}

\subsubsection{Some General Arguments}

As a result of the GR dynamics, the relativistic constraint equation
(RCE) is a restriction, which will be fulfilled at any time instant
$t$, if it is valid at the initial time instant $t=0$. Therefore it
is enough to solve RCE only at the initial time instant $t=0$.

One can use this specific feature of the RCE to simplify it,
imposing special additional conditions at the initial time instant,
which can not be fulfilled during the further evolution of the
physical system. Thus one can impose different {\em physical}
conditions on the initial state of the system under consideration.

For example, as stressed already by Misner in \cite{Misner} and a
bit later by Fock in \cite{books}, it is impossible to have a
permanent static solution for N-particle system in GR, if $N>1$.
Nevertheless, it is possible to find an initial-instant-static
solutions of the problem with any number of particles $N$. As a
rule, the initial conditions will contain some initial amount of
gravitational waves. One of the basic open problems is how to
exclude the presence of initial gravitational waves. One can expect
that for such solutions RCE will take its simplest form.

It turns out that under the conditions $\partial_t
g_{\mu\nu}|_{t=0}=0$ on the time-derivatives of the metric and some
weak additional constraint on $\partial^2_{t^2} g_{\mu\nu}|_{t=0}$
the RCE, together with other EE, yields a quasi-linear equation for
the relativistic potential of any mass distribution. This equation
can be considered as a relativistic analog to the linear equation
(\ref{Poisson}) for Newton potential in the classical theory of
gravity \cite{Misner}.

Let us consider a system of N point particles with bare masses
$M_A$, $A=1,\dots N$ at positions ${\bf r}_A(t)$. Here we suppose to
work with space-time manifold, which allows an existence of global
time $t$. The energy-momentum tensor of such system at time instant
$t$ is:
\ben T^\nu_\mu(t,{\bf r})=\sum\limits_{A=1}^N
M_A\delta_{g}^{(3)}({\bf r}-{\bf r}_A(t))\,u^{\,\,\,\nu}_{\!A}(t)
u^{}_{\!A\,\mu}(t). \hskip .5truecm \la{TmunuN}\een

According to the relativistic nonlinear superposition principle,
described in Section 1.2.4, the metric $g_{\mu\nu}(t,{\bf r})$ of
this N-particle problem is given by formula (\ref{SN}). This formula
does not yield any simple practical results. It describes in a
formal way the solution of the very complicated GR problem under
consideration and demonstrates its existence and uniqueness under
proper boundary conditions.

As we shall show in this Section, in contrast, one can introduce a
simple quasi-linear superposition principle for the relativistic
gravitational potential $\varphi({\bf r})$.

We shall consider the special case in which at the initial time
instant $t=0$ all particles are at rest and have initial positions
${\bf r}_A(0)={\bf r}_A$. Then
\ben T^\nu_\mu(0,{\bf r})=\left(\sum\limits_{A=1}^N
M_A\delta_{g}^{(3)}({\bf r}-{\bf r}_A)\right)
\,\delta^\nu_0\,\delta^0_\mu. \hskip .5truecm \la{TmunuN0}\een

According to articles \cite{Misner}, the generalization of the
covariant form of RCE (\ref{3R_point}) for N point particles at rest
{\em and} under additional conditions $\partial_t
g_{\mu\nu}|_{t=0}=0$, i.e. in the case of the N-particle {\em
instant-gravistatics}, is:
\ben {}^3\!R=16\pi \sum\limits_{A=1}^N M_A\delta_g^{(3)}({\bf
r}-{\bf r}_A).\la{3R_Npoints} \een

The proper generalization of the relation (\ref{icc}), which follows
in the {\em instant-gravistatic} case from  the EE for N-particles
reads:
\ben \int_{\mathbb{M}^{(3)}}d^3 {\bf r}\sqrt{|{}^3 g({\bf
r})|}\Delta_g\left(\sqrt{g_{tt}}\right) \psi({\bf r})=4\pi
\sum\limits_{A=1}^N m_A\psi({\bf r}_A). \la{iccN}\een
Here $m_A=M_A\sqrt{g_{tt}({\bf r}_A)}$ are the corresponding
Keplerian masses of the point particles.

A remarkable feature of the equation (\ref{iccN}) is that under
conformal Fock transformation:
\ben
g_{ij}=e^{-2\varphi}h_{ij},\,\,\,g^{ij}=e^{2\varphi}h^{ij},\,\,\,
 \sqrt{|{}^3g|}=e^{-3\varphi}\sqrt{|{}^3h|},\la{Fock}\een
one obtains the {\em quasi-linear} equation for the relativistic
potential of the N-particle problem:
\ben \partial_i \left(\sqrt{|{}^3h|}h^{ij}\partial_j\varphi
\right)=4\pi \sum\limits_{A=1}^N M_A\,
e^{\varphi_A}\delta^{(3)}({\bf r}-{\bf
r}_A),\,\,\,\hbox{where}\,\,\,\varphi_A:=\varphi({\bf r}_A).
\la{phiN}\een
Here $h_{ij}$ are functions which define the unknown metric in the
Fock conformal space of the N-particle case. In Eq. (\ref{phiN}) we
have used the substitution $\sqrt{g_{tt}}=e^\varphi$.

After the pioneering work by Lichnerowiz, one usually supposes the
metric $h_{ij}$ to be conformally flat (see in \cite{Misner}). This
leads to a well known superposition principle in the case of instant
gravistatics of N Schwarzschild black holes \cite{Misner}.

Today it is well known that the conjecture of conformal flatness of
$h_{ij}$ is too restrictive and does not allow one to obtain the
solutions of real physical problems (see the review article by Cook
in \cite{Misner} and the references therein). Unfortunately, at
present we do not know an alternative assumption, which fixes the
metric $h_{ij}$ in a physically acceptable way for the case of N
matter bodies.

\subsubsection{The Superposition Principle for the Potential $\varphi$}

Here we outline a complete different approach to the problem at
hand. It is based on a specific superposition principle in GR. This
new approach may turn to be a more physical alternative to the
conformally flat one, mention in the previous Section. The novel
superposition principle may play the role of the additional physical
requirement, needed to select the proper initial conditions for the
GR N-body-problem between all mathematically admissible and formal
initial conditions.

Having in mind the inhomogeneous quasi-linear equation (\ref{phiN}),
it seems natural to define its solution, we are looking for, by the
formula
\ben \varphi({\bf r; r_1,\dots,r_N })=-\sum_{A=1}^N{{m_A}\over
{|{\bf r-r_A}|+R_A  } } \la{DSPP_N}.\een
It follows a simple quasi-linear superposition principle for the
static relativistic potential $\varphi$. Here the Keplerian masses
are $m_A=\varrho_A M_A$ and the mass-defect ratios $\varrho_A({\bf
r_1,\dots,r_N })=\exp \varphi_A({\bf r_A; r_1,\dots,r_N })$ of the
A-th particle in presence of the other massive points can be
obtained from the self-consistency condition for the relativistic
potential
\ben \la{consistency} \varphi({{\bf r}_A; {\bf r_1},\dots,{\bf r}_N
})&=& \sum_{B=1}^N C_B \,\varphi(|{\bf r}_A-{\bf r}_B|,M_B,R_B),
\\
\varphi(|{\bf r}_A-{\bf r}_B|,M_B,R_B)&=& {{m_B}\over {|{\bf r}_A-
{\bf r}_B|+R_B } }. \een

Note that the procedure, based on the relations (\ref{DSPP_N}) and
(\ref{consistency}), represents the {\em only} way to construct a
relativistic gravitational potential $\varphi({\bf r; r_1,\dots,r_N
})$, (and corresponding metric) of a system of N point particles
with the following

{\bf Fundamental property:}
{\em For a system of N particles at finite and nonzero mutual
distances the total variety of singularities of the relativistic
potential in the whole complex domain of the variables is just a
superposition of the singularities of the relativistic potentials of
the separate matter constituents of the system.}

In other words, joining several point particles in a gravitationally
interacting system, we remain just with the singularities of all
independent particles. As a result of the integration of the
independent particles in a joint interacting system no additional
new singularities emerge, as well as no old singularities disappear
in the whole complex domain of space variables. In addition, the
singularities of the separate particles do not change their
character.

From analytical point of view one can reach such simple preservation
of the singularities only by constructing a linear combination of
the corresponding analytical functions with some constant
coefficients. The uniqueness of this construction, up to the choice
of the constant coefficients, is guaranteed by the corresponding
theorem of the complex analysis, which stays that every analytical
function is unambiguously defined by its singularities.

In the quasi-linear superposition principle (\ref{DSPP_N}) we are
using the single point particle solutions (\ref{phiGR}), denoting by
$\varrho_A^\infty=\exp\left(\varphi_A^\infty\right)$ the value of
the mass defect ratio of the A-th particle in the case
$r_{AB}:=|{\bf r}_A-{\bf r}_B|\to \infty$ for all $B\neq A$, i.e.,
in the previously considered case of a single massive point particle
in the whole universe. The constants
$C_A=\varrho_A/\varrho_A^\infty>0$ are unknown and have to be
justified.

The above consistency condition (\ref{consistency}) yields the
following basic nonlinear algebraic system of $N$ equations:
\ben \varphi_A=-\sum_{B=1}^N {{M_B\,
e^{\varphi_B}}\over{r_{AB}+R_B}}, \,\,\,\,\,A=1,\dots,N;
\la{mde}\een
for the mass defect ratios $\varrho_A=\exp\left(\varphi_A\right)$ of
the A-th particle as a member of the N-particle system in finite
space range, i.e. when all $r_{AB}<\infty$. The system (\ref{mde})
is regular one when $R_A\neq 0$ for all values of $A=0,..., N$.

In the $N$ relations (\ref{mde}) we have too many free parameters,
which have to be fixed using some proper physical assumptions.
Taking into account that the bare mass $M_A$ and the relativistic
shift $R_A=M_A\big/\left({1\over {\varrho_A^\infty}}\ln{1\over
{\varrho_A^\infty}}\right)$ are inner characteristics of the very
A-th point particle, we can suppose both of them to be constants,
whose values are independent of the N-particle configuration.

Then we see that our quasi-linear superposition principle
(\ref{DSPP_N}) leads, after all, to a definite mathematical
formulation of a novel relativistic Mach-like  principle. It states
that the Keplerian masses $m_A=m_A(r_{12},\dots,r_{N-1,N})$ of the
bodies depend on the mass distribution in the universe, in contrast
to their bare masses $M_A$, which remain independent of matter
distribution. These essentially different properties of the masses
$m_A$ and $M_A$ seem to be natural in the relativistic theory of
gravity and may have important physical consequences.

The independence of the bare particle masses $M_A$ of the system
configuration seems to be quite natural assumption. In addition we
will assume the bare mass to be an additive quantity, i.e., the
total bare mass $M$ of the composite system of particles is just the
sum of the bare masses of the constituent particles: $M=\sum_{A=1}^N
M_A$.

More speculative, from physical point of view, is the requirement of
the independence of the relativistic shifts $R_A$ from the particle
configuration. In the present article we test this assumption as a
way to restrict the number of the free parameters in the problem at
hand and study some of its consequences, leaving for future
developments its justification on a more profound physical basis.

\subsubsection{The Non-Relativistic Limit of the N-Particle Potential
$\varphi$} The restoration of the correct physical units draws an
additional light on the proposed quasi-linear superposition
principle (\ref{DSPP_N}). In physical dimension-full quantities Eq.
(\ref{DSPP_N}) acquires the form
\ben \varphi({\bf r; r_1,\dots,r_N })=\sum_{A=1}^N
\,\,{{-G^{{}^{{}_{Newton}}} M_A\, e^{\varphi_A/c^2}}\over{|{\bf
r-r_A}|-G^{{}^{{}_{Newton}}}M_Ae^{\varphi_A^\infty/c^2}\Big/\varphi_A^\infty}},
\la{DSPP_N_c} \een
and the consistency conditions (\ref{mde}) read:
\ben \varphi_A=
e^{\left(\varphi_A-\varphi_A^\infty\right)/c^2}\,\varphi_A^\infty
-\sum_{B\neq A}^N\,\, {{G^{{}^{{}_{Newton}}} M_B\,
e^{\varphi_B}}\over{r_{AB}- G^{{}^{{}_{Newton}}} M_B
e^{\varphi_B^\infty/c^2}\Big/\varphi_B^\infty}},
\,\,\,\,\,A=1,\dots,N. \la{mde_c}\een
The last formulas show directly that under the two assumptions
(\ref{varphi0_c_limits}) in the limit $c\to \infty$ one obtains
precisely the Newtonian superposition principle (\ref{DSPP}), as it
should be for the correct relativistic generalization of the
non-relativistic theory. This observation increases our confidence
in the approach, based on the superposition principle
(\ref{DSPP_N}).

To proof this statement one should take into account that:

a) $lim_{c\to \infty}\,\varphi_A^\infty=-\infty$ and  $lim_{c\to
\infty}\,\left(\varphi_A^\infty/c^2\right)=0$ -- according to the
relations (\ref{varphi0_c_limits}).

b) Using the last relations in the consistency condition
(\ref{mde_c}) one obtains easily first
\ben lim_{c\to \infty}\,\left(\varphi_A/c^2\right) &=& 0,
\la{varphi_A_c_to_infty1}\een
and then
\ben lim_{c\to \infty}\,\varphi_A &=& -\infty.
\la{varphi_A_c_to_infty2}\een

c) As a result $\varrho_A=e^{\varphi_A/c^2} \to 1$ and $m_A =
\varrho_A M_A \to M_A$ when $c\to \infty$.

\subsubsection{The Case of Continuous Mass Distribution}

The generalization of the equations (\ref{DSPP_N}) for continuous
distribution of mass is straightforward and reads:
\ben \varphi({\bf r})=-\int_{\mathbb{M}^{(3)}}d^3 {\bf r}^\prime
{{\mu({\bf r}^\prime)\, e^{\varphi({\bf r}^\prime)}}\over{ |{\bf
r}-{\bf r}^\prime|+R({\bf r}^\prime)}}.\la{mde_C}\een
Here $\mu({\bf r})$ is the density of bare mass $M$. In the case of
continuous mass distribution the relativistic shift $R$ is $R({\bf
r})=\mu({\bf r})/\chi^\infty({\bf r })$. The local density
$\chi^\infty({\bf r })$ of the quantity $x^\infty$ is the second
independent function, needed for description of the density of the
relativistic point potential in the case of continuous mass
distributions.

The relation (\ref{mde_C}) generalizes and replaces the non-relativistic
superposition principle (\ref{SPP}) for continuously distributed masses.
The last can be derived from equation (\ref{mde_C}), taking the limit
$c\to\infty$ precisely in the same way, as in the discrete case.

From mathematical point of view the relation (\ref{mde_C}) is
{\em a nonlinear and nonsingular} integral
equation for the relativistic potential $\varphi({\bf r})$:
\ben \varphi({\bf r})=\int_{\mathbb{M}^{(3)}}d^3 {\bf r}^\prime
K({\bf r},{\bf r}^\prime)e^{\varphi({\bf r}^\prime)}\la{intE}\een
with a nonsingular kernel
\ben K({\bf r},{\bf r}^\prime)=
-\,{{\mu({\bf r}^\prime)\over{|\bf
r}-{\bf r}^\prime|+R({\bf r}^\prime)}}.
\la{K}\een

For continuous distribution of identical particles with fixed
$\chi^\infty=const$ one has to put $R({\bf r})=\mu({\bf
r})/\chi^\infty$ in (\ref{K}), thus remaining with only one given
function $\mu({\bf r})\geq 0$ in the kernel $ K({\bf r},{\bf
r}^\prime)$. The function $\mu({\bf r})$ reduces to a given constant
$\mu=const$ for homogeneous mass distributions. Hence, in the last
case $R=const$, too.

\subsection{Some Basic Solutions of the Mass Defect Equation}

In this Subsection we are testing our basic assumptions, described
in the previous Sections, considering both the cases of a few point
particles, and of continuous mass distributions.

\begin{figure}[hl] \vspace{1.truecm}
 \begin{center}
         \includegraphics[totalheight=6.0cm,keepaspectratio]{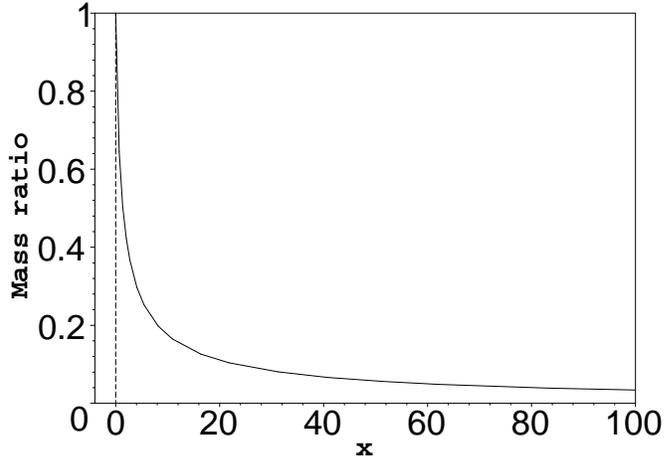}
         \vspace{-.8truecm}
        \caption{The dependence (\ref{varrho_x}) of the mass ratio
        $\varrho$ on the variable $x$.} \label{f00}
    \end{center}
\end{figure}

For this purpose is convenient to introduce the quantity
\ben x:={M\over R}={1\over \varrho}\ln{1\over \varrho}\geq 0,
\la{x}\een
which turns to play a basic role in our considerations and
demonstrates some simple properties. Making use of the Lambert
function $W(z)$ \cite{LambertW}, i.e., the solution of the equation
$$W e^W=z\,\,\, \Rightarrow\,\,\, W=W(z),$$
one obtains, solving the equation (\ref{x}) with respect to the mass
ratio $\varrho$, the basic relation (see Fig. \ref{f00})
\ben \varrho=\varrho(x)=e^{-W(x)}\la{varrho_x}.\een

Now one can write down the basic equations (\ref{mde}) in the form:
\ben \ln\left({1\over {\varrho_A}}\right)=\sum_{B=1}^N
{{x_B^\infty\, \varrho_B}\over{1+ \,r_{AB}/R_B}}. \la{Bmde}\een
We shall call this equations {\em mass defect equations} (MDE). One
immediately obtains from MDE two important consequences:

\subsubsection{A Few Particle Solutions}

1. Let us consider first the two-particle solutions (N=2) of the
MDE:

\ben \ln\left({1\over \varrho_1}\right)= x_1^\infty \varrho_1 +
 {{x_2^\infty\varrho_2}\over{1+r_{12}/R_2}}\nonumber\\
\ln\left({1\over \varrho_2}\right)=
{{x_1^\infty\varrho_1}\over{1+r_{12}/R_1}}+ x_2^\infty\varrho_2
\la{N_2}\een

For them we have two interesting limiting cases:

i) Total decay of the two-particle system, when $r_{12}\to\infty$.
Then obviously

$$\varrho_1(r_{12})\to \varrho_1^\infty,\,\,\,
\varrho_2(r_{12})\to \varrho_2^\infty,$$

as one expects.

ii) Merger of two particles $r_{12}\to 0$:
$$\varrho_{1 \cup 2}(\varrho_1^\infty,\varrho_2^\infty)=\exp\Big(
-W\left(x_1^\infty + x_2^\infty\right)\Big).$$
Note that for the quantities $x$ we obtain a simple linear
superposition:
\ben x_{1 \cup 2}=x_1^\infty+ x_2^\infty. \la{x12}\een
As a result
\ben {1\over R_{1 \cup 2}}={\mu_1\over R_2}+{\mu_2\over
R_1}=:
\left<{1\over R}\right>_{\!\!\mu},\,\,\,\,\,\mu_{1,2}=
{{M_{1,2}}\over{M_1+M_2}}\in
[0,1].\la{overR12}\een
Introducing the quantities
\ben \delta M:={{M_1-M_2}\over{M_1+M_2}} \,\,\,\hbox{and}\,\,\,
\delta R:={{R_1-R_2}\over{R_1+R_2}},\la{dRdM}\een
we obtain finally
\ben R_{1 \cup 2}=\left<{R}\right>_{\!ar}{{1-\delta
R^2}\over{1-\delta M\delta R}}. \la{R12}\een
Here $\left<{R}\right>_{\!ar}:={1\over N}\sum_{A=1}^N R_A$
denotes the arithmetic average. In the present case $N=2$.

Now we can derive easily the following basic properties of the function
$R_{1 \cup 2}=R_{1 \cup 2}\left(R_1,R_2;M_1,M_2\right)$:

1)$R_{1 \cup 2}\left(R_1,R_2;M_1,M_2\right)=
R_{1 \cup 2}\left(R_2,R_1;M_1,M_2\right)$.

2)$R_{1 \cup 2}\left(R_1,R_2;M_1,M_2\right)=
R_{1 \cup 2}\left(R_1,R_2;M_2,M_1\right)$.

3)$R_{1 \cup 2}\left(kR_1,kR_2;M_1,M_2\right)=kR_{1 \cup
4}\left(R_1,R_2;M_1,M_2\right)$, $\forall\, k>0 $.

5)$R_{1 \cup 2}\left(R_1,R_2;kM_1,kM_2\right)=R_{1 \cup
6}\left(R_1,R_2;M_1,M_2\right)$, $\forall\, k>0 $.

7)$R_{1 \cup 2}\left(R,R;M_1,M_2\right)=R$ , $\forall\,
M_1,M_2 $.

8)$R_{1 \cup 2}\left(R_1,R_2;M,M\right)={{2R_1R_2}\over{R_1+R_2}}$,
or $$R_{1 \cup 2}\left<{R}\right>_{\!ar}=\left<{R}\right>_{\!gm}^2.$$

Here $\left<{R}\right>_{\!gm}:=\sqrt{R_1R_2}$
denotes the geometric average.

9)$R_{1 \cup 2}\left(R_1,0;M_1,M_2\right)=0$.

10)$R_{1 \cup 2}\left(R_1,\infty;M_1,M_2\right)=
{{R_1}\over{\mu_1}}={{M_1+M_2}\over{x_{1 \cup 2}}}$, where $x_{1
\cup 2}=x_1$.

\vskip .3truecm

iii) In the case of {\em two identical particles}: $M_1=M_2=M,\,\,\,
R_1=R_2=R$, at finite distance $r_{12}$ one easily obtains for
$\varrho_1(r_{12})=\varrho_2(r_{12})=\varrho(r_{12})$
\ben
\varrho(r_{12})=\exp\Bigg(-W\left({{2+r_{12}/R}\over{1+r_{12}/R}}
\,x^\infty\right)\Bigg)\la{N_2_identical}\een
and
$$\varrho_{1\cup 2}(\varrho^\infty)=\exp\Bigg(
-W\left( {2\over\varrho^\infty}\ln{1\over
\varrho^\infty}\right)\Bigg).$$

\begin{figure}[hl] \vspace{.5truecm}
 \begin{center}
         \includegraphics[totalheight=6.0cm,keepaspectratio]{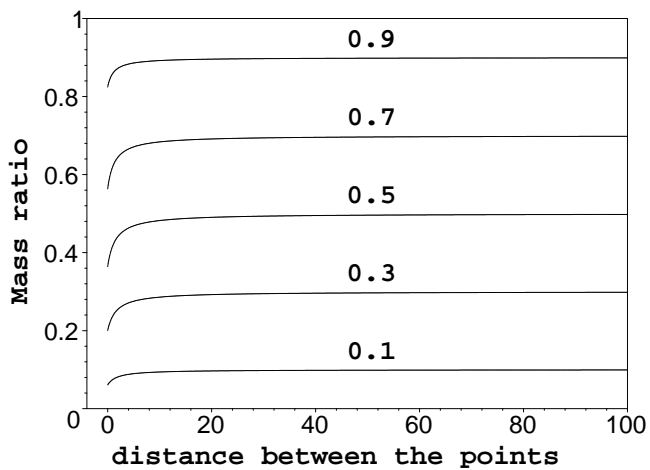}
         \vspace{-.8truecm}
        \caption{The dependence (\ref{N_2_identical}) of the mass ratio
        $\varrho(r_{12})=m(r_{12})/M$ of system of the two identical particles with
        different individual mass ratios $\varrho^\infty=0.1,\dots, 0.9;$
        on the distance $r_{12}$ between them.
        The distance $r_{12}$ is shown in units of $R$.} \label{f01}
    \end{center}
\end{figure}

The small variation of the mass ratio of each of the two identical
particles in the system with the change of the distance between the
particles is shown in Fig. {\ref{f01}}. As seen, at distances
$r_{12}\gg R$ the measurable Keplerian mass of each particle is
practically constant, since the value of $\varrho$ is almost
constant.

\begin{figure}[hl] \vspace{.5truecm}
 \begin{center}
         \includegraphics[totalheight=6.0cm,keepaspectratio]{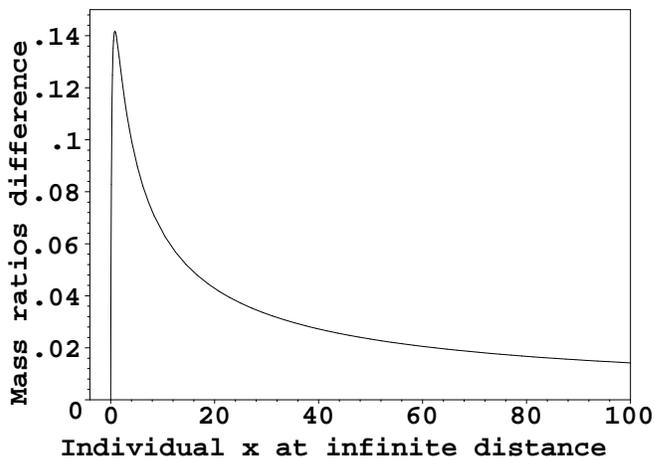}
         \vspace{-.5truecm}
        \caption{The dependence (\ref{Delta_vrho}) of the mass ratio
        difference $\Delta \varrho(x^\infty)$ on the individual value of  $x^\infty$
        for the two identical particles.} \label{f02}
    \end{center}
\end{figure}

The formula
\ben
\Delta \varrho(x^\infty)=\varrho(0)-\varrho(\infty)=
e^{-W(x^\infty)}-e^{-W(2\,x^\infty)}
\la{Delta_vrho}\een
shows that the variation of of the mass ratio of each particle
$\varrho(r_{12})$ is not bigger then $\approx 0.1408$, when the
distance between them  varies from zero to infinity:
$r_{12}\in(0,\infty)$, see Fig. \ref{f02}.
\begin{figure}[hl] \vspace{.5truecm}
\begin{center}
        \includegraphics[totalheight=6.0cm,keepaspectratio]{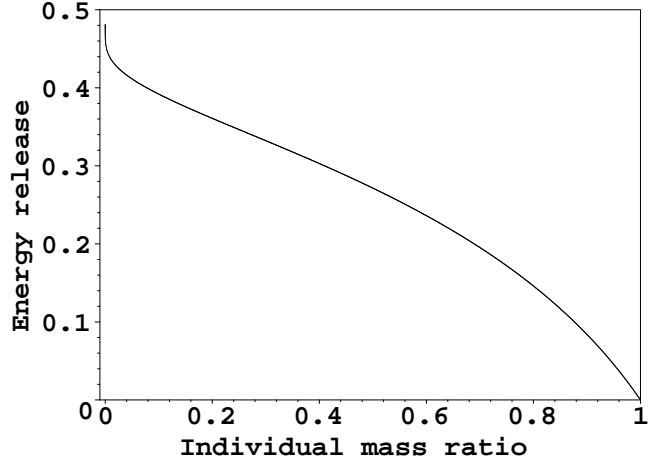}
        \vspace{-.5truecm}
       \caption{The dependence (\ref{EnEf}) of the energy release (in units $2Mc^2$)
       on the individual mass ratio $\varrho^\infty$ of the two identical particles.}
       \label{f03}
   \end{center}
\end{figure}

The energetic efficiency of the process of gravitational merger of
two identical particles is described by the quantity:
\ben
{{\Delta E_G}\over{2Mc^2}}=1-\varrho_{1\cup 2}(\varrho^\infty)/\varrho^\infty.
\la{EnEf}\een

Here $\Delta E_G$ is the energy release in the gravitational
collapse of the pair of the two identical point particles from
infinite distance to their merger.

2. For three identical particles at the vortices of equilateral triangle
$r_{AB}=r$ for all $A,B=1, 2, 3; A\neq B$ one obtains:
\ben
\varrho(r)=\exp\Bigg(-W\left({{3+r/R}\over{1+r/R}}\,x^\infty\right)\Bigg).
\la{N_3_identical}\een

3. For four identical particles at the vortices of equilateral
tetrahedron $r_{AB}=r$ for all $A,B=1, 2, 4; A\neq B$ the result is:
\ben
\varrho(r)=\exp\Bigg(-W\left({{4+r/R}\over{1+r/R}}\,x^\infty\right)\Bigg).
\la{N_4_identical}\een

\subsubsection{Some Basic Properties of the N-Particles Solutions}

Consider now a system of $N$ particles:

i) If the system of N point particles decays, i.e., when
$r_{AB}\to\infty$ for all $B\neq A$ and we remain with only one
such particle in the whole universe, then $x_A\to x_A^\infty$, as
it should be.

\vskip .3truecm

\noindent ii) In the opposite case, when all $r_{AB}\to 0$, i.e. when
the system N particle collapses and they merge into one composite particle,
from MDE one obtains $\varrho_A=\varrho_B=const=\varrho_{1\cup\dots \cup N}$ for all A, B.
For the resulting mass defect ratio one has
\ben \varrho_{1\cup\dots \cup N}=\exp\big(-W\left(x_{1\cup\dots \cup
N}\right)\!\big)=\exp\left(-W\left(\sum_{A=1}^N
x_A\right)\right)\la{vrho_N}\een
and the simple linear superposition:
\ben x_{1\cup\dots \cup N}=\sum_{A=1}^N x_A.\la{x_N}\een

\noindent iii) For an aggregate of N fused
particles of total bare mass $M$ we obtain:
\ben 1/R_{1\cup\dots \cup N}=\sum_{A=1}^N \mu_A/R_A =:
\left<{1/R}\right>_{\!\mu},\\ \mu_A={{M_A}\over{M}}\in(0,1),
\,\,\,\sum_{A=1}^N\mu_A=1. \nonumber
\la{M_N_R_N}\een
\noindent iv) Using the asymptotic of function W(z) one easily
obtains for the merger of N identical point particles, each of bare mass $M$
and relativistic shift $R$
\ben \varrho_{1\cup\dots \cup N}=\exp\left(-W\left(N
{{M}\over{R}}\right)\right)\sim {{\ln
\left(N{{M}\over{R}}\right)}\over{ \left(N{{M}\over{R}}\right)}}
\,\,\,\hbox{-- when}\,\,\,N\to\infty \la{N_identical}\een
and
\ben R_{1\cup\dots \cup N}=R,\,\,\,\,\,\forall\, N.\la{R_N}\een
Here $R$ is the relativistic shift of the separate particle.

These formulas show that the accumulation of particles during the merger
of N particles leads to increase of the mass defect,
since $\varrho_{1\cup\dots \cup N}\to 0$
when $N\to \infty$ -- much like the situation in nuclear physics,
as it should be from physical point of view.

\subsubsection{Some Solutions of the Mass Defect Equation
for Continuous Distributions}

To acquire  some experience working with solutions of the MDE in the
case of continuous mass distributions, we will consider in this
subsection several simple examples of such distributions of
identical particles with a constant density  $\mu$ and a simple
geometry:

1. First we consider one-dimensional continuous mass distribution on
a homogeneous circle of diameter $d$, and linear mass density
$\mu_1$, made of identical particles.

From a symmetry reasons at the very circle one has $\varphi=const$.
Then one has to calculate a simple integral in the rhs of Eq. (\ref{mde_C}).
Thus one obtains for the mass ratio the following expression:
\ben \varrho(d)= \exp\left(-W\Bigg({{\mu_1}\over{\sqrt{1-R^2/d^2}}}
\ln{{1+\sqrt{1-R^2/d^2}}\over{1-\sqrt{1-R^2/d^2}}}\Bigg)\right).
\la{circle} \een

\begin{figure}\vspace{.5truecm}
    \begin{center}
        \includegraphics[totalheight=6.0cm,keepaspectratio]{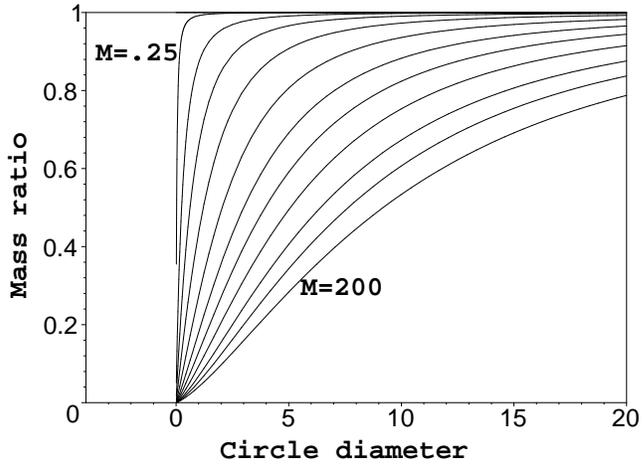}
        \caption{The dependence of the mass ratio
        $\varrho_{circle}=m/M$ on the diameter $d$
         of circles with different fixed total bare masses $M$.
         The diameter $d$ is shown in units of $R$.} \label{f1}
    \end{center}
\end{figure}

Suppose that the total bare mass of the circle is $M$. Then
$\mu_1=M/\pi d$. Replacing $\mu_1$ in the relation (\ref{circle})
with this value, we obtain the result, shown in the Fig. \ref{f1}.
As seen, $\varrho(0)=0$ and when the masses of the circle
are dispersed at bigger distances,
increasing its diameter, the mass ratio increases
and goes to 1 for $d\to\infty$,
as it should be from physical point of view.

\begin{figure}[hl] \vspace{1.truecm}
 \begin{center}
         \includegraphics[totalheight=6.0cm,keepaspectratio]{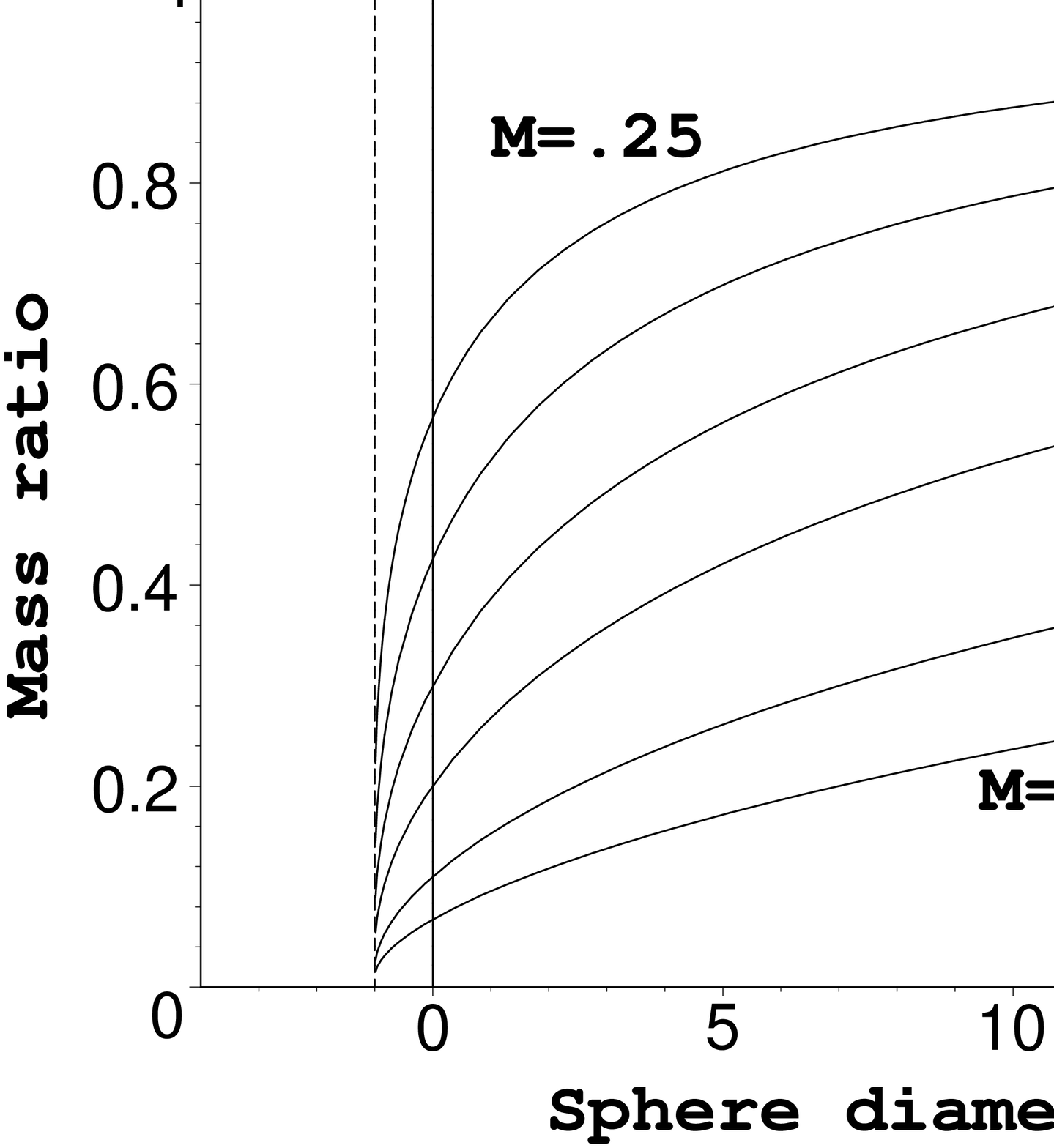}
        \caption{The dependence of the mass ratio
        $\varrho_{sphere}=m/M$ on the diameter $d$
        of spheres with different fixed total bare masses $M$.
        The diameter $d$ is shown in units of $R$.} \label{f2}
    \end{center}
\end{figure}

2. Our second example is a two-dimensional continuous mass
distribution on a homogeneous sphere of diameter $d$
and surface mass density $\mu_2=4 M/\pi d^2$, made of identical particles.
In this case one obtains in a similar way for the mass ratio:
\ben \varrho(d)=\exp\Bigg(-W\Big({{8M}\over d}\,\big(1-{R\over d}
\ln\left(1+d/R\right)\big)\Big)\Bigg).\la{sphere}\een
As seen in Fig. \ref{f2}, when the masses of the sphere are
dispersed at bigger distances, increasing the diameter, the mass
ratio increases and goes to 1, when $d\to\infty$,
as it should be from physical point of view.

In addition we see that in this example the mass ratio goes
to zero for the non-physical value of sphere diameter $d=-R$.
In contrast to the previous example, now one has $\varrho(0)=e^{-W(4M/R)}>0$.
This new property reflects the peculiar
geometry of the spheres around the point sources:
the limit of the sphere surface area remains finite
when the sphere diameter goes to zero.

3. For a homogeneous three-dimensional ball of radius $r_*$ one easily obtains the
{\em one dimensional} integral equation:
\ben \varphi(r)=\int_0^{r_{*}}dr^\prime
k(r,r^\prime)e^{\varphi(r^\prime)}\la{ball}\een with kernel
\ben k(r,r^\prime)=-{{2\pi\mu}\over{r}}r^\prime\Bigg(r+r^\prime
-|r-r^\prime| -R\ln\left({{r+r^\prime+R}
\over{|r-r^\prime|+R}}\right)\Bigg).\la{kernel_ball}\een
Let us consider the limiting case of Eq. (\ref{ball}) with $R=0$,
and let us denote $\varphi_{[0]}=\varphi|_{R=0}$. Then
we obtain the following Debye-H\"uckel like boundary
problem:
\ben
{{d^2\varphi_{[0]}}\over{d\xi^2}}+{{2}\over{\xi}}{{d\varphi_{[0]}}\over{d\xi}}=
e^{\varphi_{[0]}},
\,\,\,\,\,\,\xi=\sqrt{4\pi\mu}\,r, \nonumber\\
{{d\varphi_{[0]}}\over{d\xi}}(0)=0,\,\,\,\xi_*
{{d\varphi_{[0]}}\over{d\xi}}(\xi_*)+\varphi(\xi_*)=0.
\la{DH_like}\een
This problem has no auxiliary parameters and defines
an {\em universal} function $\varphi_{[0]}(\xi,\xi_*)$. The solution gives the
zero order term in the exact solution of the integral equation
(\ref{ball}): $\varphi(r)=\varphi_{[0]}(\sqrt{4\pi\mu}\,r)+ {\cal O}(R)$
and obviously describes the gravitational screening of the bare mass
in the present simple case of three-dimensional continuous mass distribution.

It is not hard to find the numerical solutions of the problem (\ref{DH_like}).

\section{Some Concluding Remarks}
There exist at least three different approaches to the relation
between physics and geometry:

A) The classical physics considers the space-time continuum only
as arena for the struggles of fields and particles. "These
entities are foreign to geometry. They must be added to geometry
to permit any physics" (see, for example, Misner and Wheeler in
\cite{Misner}). In this approach the geometry is complete
independent of physics.

B) According to original Einstein's idea, the geometry of
space-time is determined by the matter sources of gravity. Masses
and non-gravitational fields are of non-geometrical origin. To
some extend these sources of gravity are independent of geometry
entities. In their presence the space-time continuum acquires the
geometrical properties of curved 4D pseudo-Rieamannian manifold.
Its geometry determines the motion and evolution of the very
matter. Hence, in this approach we have a complicated interplay
between geometry and physics, based on nonlinear differential
equations.

C) There is a third way, pioneered by Einstein and Rosen and
developed in pure "geometrodynamics" by Wheeler and others. In
this approach one may think that "There is nothing in the world
except empty curved space. Matter, charge, electromagnetism and
other fields are only manifestation of the bending of space.
Physics is geometry.", see Misner and Wheeler in \cite{Misner}. In
this approach we have "mass without mass, charge without charge",
etc.  At present the most well known hypothetical notions, created
in the framework of this approach are the "black holes" and the
"wormholes" of different type in space-time. They are consequences
of the assumption to have "{\em everywhere empty} curved space",
see Wheeler in \cite{Misner}.

In the present article we accept the {\em original} Einstein's
idea, described in point B, about the relation between mater and
geometry of space-time.  Here we develop the mathematical
realization of this idea considering matter {\em point} particles
as sources of gravity in GR. We have reached the following basic
results:

1. We have studied a new, two parameter class of solutions of
Einstein equations. These static spherically symmetric solutions
describe the gravitational field of massive point particle with bare
mass $M>0$ and Keplerian mass $m$ ($0<m < M$). The difference
between these masses, or their ratio $\varrho=m/M\in (0,1)$, defines
the gravitational mass defect of the point particle. Such mass
defect was not considered and studied until now, because for the
standard Hilbert form (\ref{Hilbert}) of the Schwarzschild solution
"the bare rest-mass density is never even introduced" \cite{ADM}.

2. The new solutions form a two parameter family of metrics on
singular manifolds $\mathbb{M}^{(1,3)}\{g_{\mu\nu}\}$, described in
details in the present article, as well as in \cite{PF, PFSD}.

3. We have shown the principal role of the massive point source of
gravity. Its presence offers a natural cutting for the {\em
physical} values of the luminosity variable $\rho\in [\rho_0,
\infty)$, where $\rho_0> \rho_G$. This happens because the infinite
mass density of the matter point changes drastically the geometry of
the space-time around it. This phenomenon is in sharp contrast to
the situation in geometrodynamics, where luminosity variable may
have an arbitrary close to zero value around the center of the black
holes.

A geometry of space-time with $\rho_0 \equiv \rho_G>0$ was
discovered at first in the original article by Schwarzschild
\cite{Schwarzschild}. According to Eq. (\ref{rho}), such limiting
value of the luminosity variable corresponds to zero value
$\varrho=0$ of mass defect ratio. For a finite value of $m$ this is
possible only if $M=\infty$. In this sense our work is a proper
extension of the Schwarzschild one to the physically and
mathematically admissible values of the mass defect ratio
$\varrho\in (0,1)$.

4. The existing attempts to describe the point matter source using
Schwarzschild solution in Hilbert gauge  (\ref{Hilbert}) do not take
into account the mass defect and thus fail to present a point
idealization of the real relativistic objects.

5. In full accord with Dirac's suggestion \cite{Dirac} our cutting
of the domain of luminosity variable places the event horizon in the
nonphysical domain of the variables. This effect is well known from
the solutions of Einstein equations with massive matter sources of
finite dimension \cite{books}.

6. The mathematical and the physical properties of the new solutions
are essentially different in comparison with the well known other
spherically symmetric static solutions to the Einstein equations.
All of the new solutions have a strong singularity at the center of
the symmetry, which is surrounded by empty space. They describe the
single point particle sources of gravity in GR. The previously known
static spherically symmetric solutions were often erroneously
considered as a solutions, which describe a point mass, but this is
not the case.

7. It is clear that our solutions in generalized functions define in
mathematical sense the fundamental solutions of the quasi-linear
Einstein equations. These solutions are complete analogous to the
fundamental solutions of Poisson equation in Newton theory of
gravity. Thus the problem, formulated by Feynman in \cite{Feynman}
is solved.

8. A proper quasi-linear superposition principle for initial
conditions of Einstein equations exist. It leads to a new theory of
the relativistic gravitational defect of mass, illustrated in short
in the present article. Our study shows that the hypotheses, used in
this approach to the superposition principle in GR lead to
physically reasonable consequences. One has to put on a more
profound basis these hypotheses and to compare their consequences
with the physical reality.

9. Our results show that it may turn to be possible to transform the
original Einstein relativistic theory of gravity, without change of
its dynamical equations, into a normal physical theory with basic
properties, which are intrinsic to the other branches of physics.
This may permit us to get out of the way some of the specific
scientific fictions, which are widespread at present and are thought
to be an unavoidable consequences of GR.

Especially, the correct theory of N matter particles systems seems
to lead to the space-times with Euclidean topology, global time and
standard physical causality. Hopefully, these important properties
may transmute the now-days form of GR into a new version of the
relativistic theory of gravity, compatible with the standard
relativistic quantum mechanics of particles, much like the theories
in the other branches of physics, and despite of the curvature of
the space-time.

10. Our basic conclusion is that we urgently need critical
experiments and observations, which can help us to choose the
version of Einstein relativistic theory of gravity, compatible with
the physical reality. Some idea of such type of experiments was
recently proposed and discussed in \cite{Fiziev2005}.

\vskip .8truecm
{\em \bf Acknowledgments}
\vskip .3truecm

This article was supported by Bulgarian the Scientific Found,
Contract /2006, by the Scientific Found of Sofia University,
Contract 70/2006 and by its Foundation "Theoretical and
Computational Physics and Astrophysics".

\end{document}